\documentclass[aps,preprint,preprintnumbers,amsmath,amssymb]{revtex4}
\usepackage{mathrsfs}
\usepackage{amssymb}
\usepackage{graphicx}
\usepackage{dcolumn}
\usepackage{bm}
\usepackage{color}
\usepackage[breaklinks=true,colorlinks=true,linkcolor=blue,urlcolor=blue,citecolor=blue]{hyperref}
\UseRawInputEncoding   
\usepackage{soul}
\usepackage{ulem}
\usepackage{ragged2e}
\makeatletter
\newcommand{\rmnum}[1]{\romannumeral #1}
\newcommand{\Rmnum}[1]{\expandafter\@slowromancap\romannumeral #1@}
\newcommand{\lyk}[1]{\textcolor{red}{#1}}

\begin{document}

\title{Abnormal planar Hall effect and disentanglement of incoherent and coherent transport in a Kondo lattice}

\author{Shuo Zou$^{1}$, Hai Zeng$^{1}$, Zhuo Wang$^{1}$, Guohao Dong$^2$, Xiaodong Guo$^{1}$, Fangjun Lu$^{1}$, Zengwei Zhu$^{1}$, Youguo Shi$^{2,3}$\footnote[1]{Electronic address: ygshi@iphy.ac.cn}, Yi-feng Yang$^{2,3}$\footnote[2]{Electronic address: yifeng@iphy.ac.cn} and Yongkang Luo$^{1}$\footnote[3]{Electronic address: mpzslyk@gmail.com}}

\address{$^1$Wuhan National High Magnetic Field Center and School of Physics, Huazhong University of Science and Technology, Wuhan 430074, China;}
\address{$^2$Beijing National Laboratory for Condensed Matter Physics, Institute of Physics, Chinese Academy of Sciences, Beijing 100190, China;}
\address{$^3$University of Chinese Academy of Sciences, Beijing 100049, China.}

\date{\today}


\maketitle

\textbf{
The nature of localized-itinerant transition in Kondo lattice systems remains a mystery despite intensive investigations in past decades. While it is often identified from the coherent peak in magnetic resistivity, recent angle-resolved photoemission spectroscopy and ultrafast optical spectroscopy revealed a precursor incoherent region with band bending and hybridization fluctuations. This raises the question of how the coherent heavy-electron state is developed from an incoherent background of fluctuating localized moments and then established at sufficiently low temperatures. Here, on the example of the quasi-one-dimensional Kondo lattice compound CeCo$_2$Ga$_8$, we show that planar Hall effect and planar anisotropic magnetoresistance measurements provide an effective way to disentangle the incoherent Kondo scattering contribution and the coherent heavy-electron contribution, and a multi-stage process is directly visualized with lowering temperature by their distinct angle-dependent patterns in magneto-transport. Our idea may be extended to other measurements and thereby opens up a pathway for systematically investigating the fundamental physics of Kondo lattice coherence.
}\\

\begin{figure}[!htp]
\hspace*{-0pt}
\vspace*{-0pt}
\includegraphics[width=10cm]{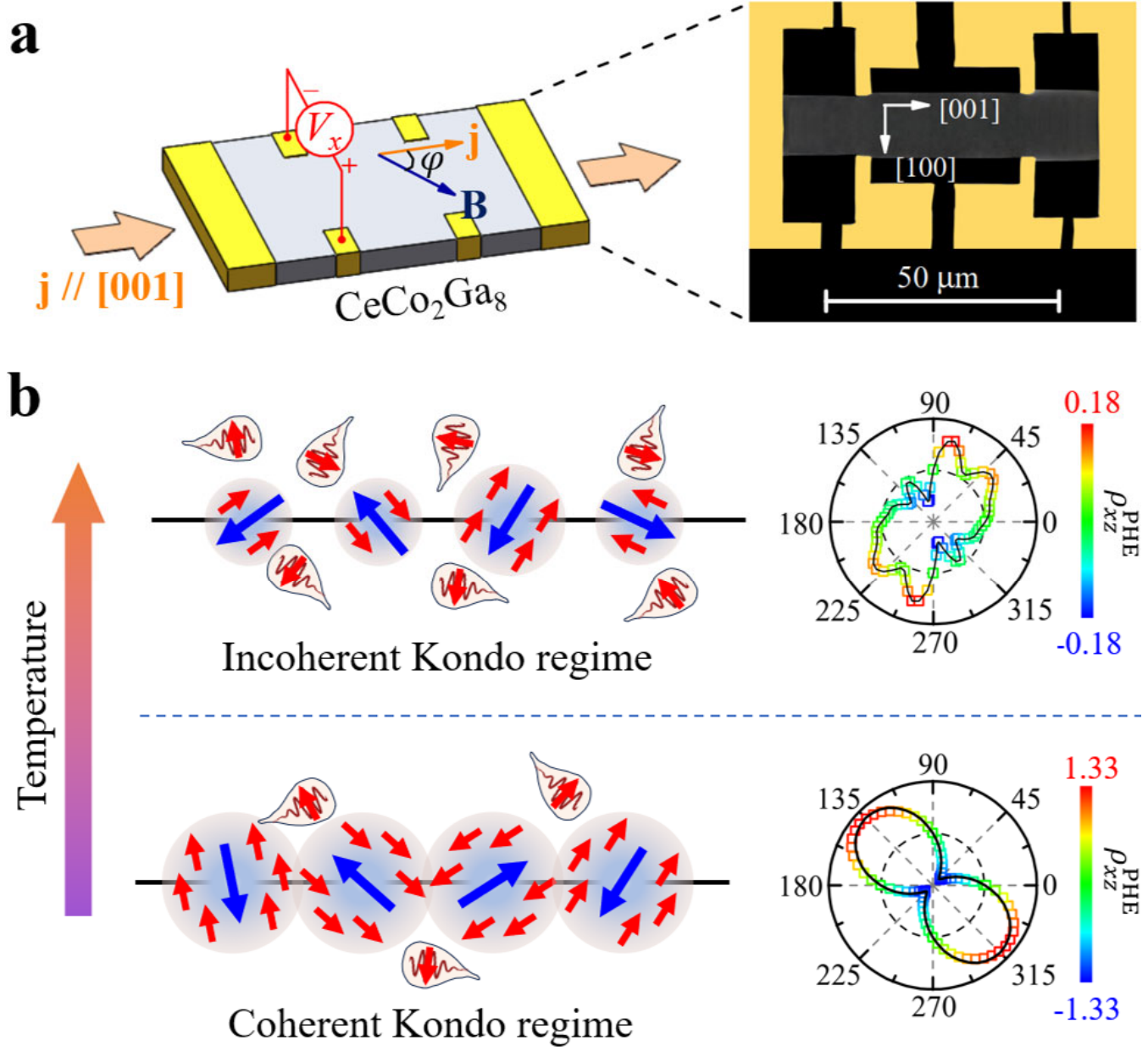}
\label{Graphical_Abstract}
\end{figure}
\vspace*{-10pt}
\textbf{Graphical abstract.} This work studies the localized-itinerant transition in Kondo lattice systems, and opens up a pathway to investigate the puzzling physics of Kondo lattice coherence. In particular, the authors demonstrate that the incoherent Kondo scattering contribution and the coherent heavy-electron contribution can be told apart by performing planar Hall effect and planar anisotropic magnetoresistance measurements.  \\

\textbf{INTRODUCTION}

The transformation of localized-itinerant electronic states is one of the key ingredients in condensed matter physics, bridging a commonality among a variety of prototypical strongly correlated electronic systems including but not limited to the high-$T_c$ superconductors, heavy-fermion metals, irradiates, organic superconductors, and twisted-bilayer graphene \cite{BrandowB-JAC1992,ImadaM-RMP1998,SiQ-localQCP,Coleman-QCP2005,Pepin-Mott2008,KimBJ-Sr2IrO4PRL2008,OikeH-PRL114_067002(2015),CaoY-MAGNature556_80_2018}. In heavy-fermion systems, such a transformation is usually connected to the Kondo hybridization between itinerant conduction ($c$) and localized $f$ electrons (viz. $c$-$f$ hybridization), the contribution of delocalized $f$ electrons to Fermi surface (FS), and hence the emergence of composite heavy-electron state. A classic description is that $c$-$f$ hybridization is fulfilled below coherence temperature $T^*$, and this scenario was well supported by traditional transport measurements in a series of Ce-based Kondo lattice compounds \cite{AlievFG-CeCu2Si2JLTP1984,HeggerH-CeRhIn5PRL2000,NakatsujiS-CeCoIn5PRL2002,LuoY-CeNiAsONM2014}. Spectroscopically, however, recent angle-resolved photoemission spectroscopy (ARPES) and ultrafast optical spectroscopy measurements revealed that hybridization already appears at much higher temperature $T_K^{on}$ and no remarkable change is detectable across $T^*$ \cite{ChenQY-PRB2017,KoitzschA-PRB2008,LiuY-PRL2020,JangaS-CeCoIn5PNAS2020}. These contradicting results imply that a precursor-like state where a short-range \textit{incoherent} hybridization regime may exist above $T^*$. To reconcile this dichotomy and to unveil the intrinsic features in this special regime as well, extensive theoretical and experimental works are both highly demanded.

Angular dependent magnetotransport properties provide a useful probe to Fermi surface topology and scattering process. In this work, we systematically investigated planar Hall effect (PHE) and planar anisotropic magnetoresistance (PAMR) on a quasi-one-dimensional (quasi-1D) Kondo lattice compound CeCo$_2$Ga$_8$. The schematic of PHE and PAMR is depicted in Fig.~1(a). Unlike ordinary Hall effect that is caused by an out-of-plane magnetic field, PHE represents an in-plane-field-rotation-induced modulation of the transverse resistivity ($\rho_{xz}^\text{PHE}$). A classic origin of PHE arises from anisotropic longitudingal resistivity ($\rho_{zz}$) under field, and can be described by the empirical formula \cite{TangHX-PRL2003,ZhongJ-CPB2023,LiL-npjQM2023}:
\begin{subequations}
\begin{align}
\rho_{zz}(\varphi)&=\rho_{\perp}-\Delta\rho \cos^2\varphi, \label{Eq.1a}\\
\rho_{xz}^\text{PHE}(\varphi)&=\Delta\rho\sin\varphi\cos\varphi, \label{Eq.1b}
\end{align}
\end{subequations}
where $\Delta\rho\equiv\rho_{\perp}-\rho_{\parallel}$ stands for the resistivity anisotropy, and $\rho_{\parallel}$ ($\rho_{\perp}$) is the longitudinal resistivity when $\mathbf{B}$ is parallel (perpendicular) with electrical current $\mathbf{j}$. Since magnetoresistance probes deviations of FS from sphericity or the variance of local Hall mobility around the FS \cite{PippardAB-MRinMetal,HarrisJM-PRL1995}, hopefully, studies of PHE and PAMR will uncover more details in electronic structure and scattering mechanism.

Herein, we report abnormal oscillation patterns in PHE of CeCo$_2$Ga$_8$ with higher harmonics below $T_K^{on}\sim100$ K, whose magnitudes first increase with lowering temperature and then rapidly diminish below the coherence temperature $T^*\approx 12$ K. Simultaneously, another two-fold oscillation mode emerges below $T^*$ and gradually dominates the PHE. At sufficiently low temperatures ($T\ll T^*$), the two-fold oscillatory PHE is recovered but with a phase switch of about $180^{\circ}$ compared to those from conduction bands near $T_K^{on}$, which indicates that the emergent mode has a different origin than conduction bands. Comparison with other measurements attributes the abnormal oscillation patterns to the incoherent Kondo scattering and the low-temperature emergent mode to the coherent heavy-electron state. Our work thus provides a way to disentangle incoherent and coherent contributions in transport measurements, and presents unprecedented transport evidence for the existence of a precursor state with dynamic hybridization. \\

\textbf{RESULTS}

\textbf{PAMR and PHE of CeCo$_2$Ga$_8$}

The compound CeCo$_2$Ga$_8$ crystallizes in the YbCo$_2$Al$_8$-type orthorhombic structure (Pbam, No. 55) \cite{Koterlin-Ce128}, as shown in Fig.~1(b). The cerium atoms form individual chains along [001], and these chains are well isolated by the CoGa$_9$ cages in the $\mathbf{ab}$ plane, rendering it a candidate of quasi-1D Kondo lattice (namely Kondo chain) \cite{WangL-CeCo2Ga8}, and this was supported by our earlier resistivity measurements which indicated that Kondo coherence develops only for electric current parallel to the chains ($\rho_{c}$), while the profiles of $\rho_{a}$ and $\rho_{b}$ remain incoherent down to 2 K \cite{ChengK-CeCo2Ga8Q1D,ChengK-CeCo2Ga8Stress}. Magnetic susceptibility, specific heat, and $\mu$SR measurements manifested that it does not exhibit any trace of long-range magnetic ordering down to 0.07 K \cite{WangL-CeCo2Ga8,Bhattacharyya-CeCo2Ga8uSR}, but rather, remarkable non-Fermi-liquid behavior characterized by linear resistivity [$\rho_c(T) \propto T$] and logarithmic specific heat [$C/T\propto-\log T$] appears at low temperature, implying in the vicinity of a quantum critical point.

To carry out the PHE and PAMR experiments, electrical contacts were made in a Hall-bar geometry on a piece of CeCo$_2$Ga$_8$ single crystal whose orientation was verified by X-ray diffraction [Fig.~S1 in \textbf{Supplementary Information} (\textbf{SI})]. To eliminate spurious contamination between the longitudinal ($\rho_{zz}$) and transverse ($\rho_{xz}^\text{PHE}$) voltages, focused-ion-beam (FIB) technique was exploited to make the microstructured device, as shown in Fig.~1(a). In our experiments, the electrical current ($\mathbf{j}$) flows along [001], whereas the magnetic field ($\mathbf{B}$) was rotated within the $\mathbf{ac}$ plane. The angle spanned by $\mathbf{j}$ and $\mathbf{B}$ is denoted by $\varphi$. The temperature dependence of $\rho_{zz}$ in the absence of $\mathbf{B}$ is shown in Fig.~1(c). It decreases upon cooling from room temperature as a metallic behavior, but turns up near $T_{K}^{on}\sim 100$ K where the incoherent Kondo scattering sets in. A broad peak is observed in $\rho_{zz}(T)$ around $T^*\sim 12$ K below which a coherent heavy-electron state is believed to emerge. $-\log T$ behavior featuring single-impurity Kondo effect is seen between $T_K^{on}$ and $T^*$, as expected. The $T^*$ in this sample is relatively lower than in our previous works ($\sim 17$ K) \cite{WangL-CeCo2Ga8,ChengK-CeCo2Ga8Q1D,ChengK-CeCo2Ga8Stress}, but it still falls within the normal range of 10-20 K observed in the majority of samples. It should also be noted that FIB treatment seems to have no essential influence on $T^{*}$, seeing Fig.~S2 in \textbf{SI}. Below $T^*$, a phenomenological two-fluid theory has predicted the coexistence of incoherent Kondo scattering and coherent heavy-electron state, but experimental evidences for their respective contributions to the transport properties have not been available.

Turning now to the PHE and PAMR of CeCo$_2$Ga$_8$. The $\varphi$ dependent $\rho_{xz}^\text{PHE}$ and $\rho_{zz}$ measured under 5 T and 1.6 K (well below $T^*$) are presented in Figs.~1(d) and (e), respectively. Two characteristic features commonly seen in planar Hall measurements are also evident here. First, both $\rho_{xz}^\text{PHE}$ and $\rho_{zz}$ are sinusoidal on the angle $\varphi$, exhibiting two-fold oscillations and obeying well to Eq.~(1). Second, a 45 $^{\circ}$ phase shift is present between $\rho_{xz}^\text{PHE}$ and $\rho_{zz}$, which is better seen in the polar-coordinate plots in Fig.~1(f-g). These features are akin to the PHE and PAMR reported previously in other compounds, many of which were topological semimetals, such as Cd$_3$As$_2$ \cite{LiH-PRB2018}, NiTe$_2$ \cite{LiuQQ-PRB2019}. Since PHE usually originates from anisotropic magnetoresistance (i.e. $\Delta\rho$), generally speaking, weak PHE can be a common feature in systems with anisotropic Fermi surface or scattering \cite{YinG-MnTe}. However, in some specific circumstances, the PHE can be further amplified, such as the presence of chiral anomaly in topological Dirac / Weyl semimetals \cite{NielsenHB-PhysLettB1983,SonDT-PRB2013,XiongJ-Science2015,LuoY-SciRept2016} or spin-momentum locked surface state in topological insulators \cite{TaskinAA-NC2017,WuB-APL2018,ZhouL-PRB2019}. In our case, anisotropic magnetoresistance is not surprising considering its quasi-1D crystalline structure. First-principles calculations also pointed to flat FSs for the itinerant $f$-electron bands \cite{WangL-CeCo2Ga8}, it then is reasonable that the magnetoresistance can be highly anisotropic. Besides, our earlier magnetic susceptibility measurements and crystalline electric field (CEF) analysis manifested $\mathbf{c}$ as the magnetic easy axis \cite{ChengK-CeCo2Ga8Q1D}, which may cause anisotropic spin scatterings. All these may contribute to the observation of PHE in CeCo$_2$Ga$_8$.\\

\textbf{Abnormal PHE in CeCo$_2$Ga$_8$}

To further look into the PHE and PAMR of CeCo$_2$Ga$_8$, we also measured $\rho_{xz}^\text{PHE}(\varphi)$ and $\rho_{zz}(\varphi)$ under various fields, and the results (at 1.6 K) are displayed in Fig.~2. Generally, the oscillating amplitude of both $\rho_{zz}(\varphi)$ and $\rho_{xz}^\text{PHE}(\varphi)$ increases as field strengthens. A major feature for the longitudinal channel is that the phase of the two-fold oscillation switches by 180$^\circ$ when magnetic field exceeds 9 T, and meanwhile some sub-feature appears around $\varphi=0$ and 180 $^\circ$ ($\mathbf{B}\parallel$ [001]). This motivates us to look also into the field dependent magnetoresistance [MR$\equiv\frac{\rho_{zz}(B)-\rho_{zz}(0)}{\rho_{zz}(0)}\times100\%$] for $\mathbf{B}\parallel$ [100] and [001], seeing Fig.~2(b). A prominent feature is that the two MR curves cross near 11.8 T, meaning $\Delta\rho$ changes sign at this field. We infer that such a sign change in $\Delta\rho(B)$ is the reason for the phase shift in the PAMR curve between 9 and 13 T. Meanwhile, we also notice that while MR for $\mathbf{B}\parallel[100]$ is negative and decreases monotonically all through the field range 0-15 T, the MR for $\mathbf{B}\parallel[001]$ increases with field and reaches a maximum $\sim 1\%$ near 6 T before it turns down and becomes negative for field larger than 10.5 T. A low-field positive MR is common for Kondo lattice compounds in the coherent regime, e.g. CeAl$_3$ \cite{RoeslerGM-PRB1992}, Ce(Co,Rh)In$_5$ \cite{ChristiansonAD-PRB2002,MalinowskiA-PRB2005}, YbNi$_2$B$_2$C \cite{YatskarA-PRB1999}, etc. The $B^2$ behavior of MR shown in Fig.~2(b) for both $\mathbf{B}\parallel$ [100] and [001] ($B>6$ T) also resemble that of CeCoIn$_5$ in the coherent regime \cite{MalinowskiA-PRB2005}. According to the phenomenological two-fluid model for heavy-fermion systems \cite{Nakatsuji-PRL2004,Yang-PRL2008}, $f$ electronic state below $T^*$ is a mixture of the localized $f$ electron (Kondo impurity fluid) and itinerant composite heavy-fermion fluid, and as $T$ decreases, the weight of the latter enhances. Based on this framework and taking also account of Matthiessen's rule, we infer that such a non-monotonic MR behavior is likely a consequence of the competition between the Kondo impurity liquid resistivity that decreases under field and the heavy Fermi liquid resistivity that behaves in an opposite trend. Note that for the former, the unscreened Ce-$4f$ electrons act as fluctuated local moments, causing incoherent Kondo scattering of conduction electrons; the reduction of spin-disorder scattering by magnetic field gives rise to the negative MR \cite{YatskarA-PRB1999}.

Getting back to PHE. Figure 2(c) shows PHE under diverse magnetic field strengths. It is interesting to note that although the phase of PAMR get reversed between 9-13 T, the phase of PHE remains unchanged, and its shape is not distorted, either. For all the fields measured 0-15 T, the curves of PHE display nice sinusoidal shapes. The amplitude of PHE increases monotonically with field and a tiny saturation trend is visible when field surpasses 13 T, seeing Fig.~2(d). Furthermore, unlike what was reported in other examples, there is no linear relationship between the amplitudes of PHE and PAMR [cf. inset to Fig.~2(d)]. An intuitive expectation is that in compensated metals the amplitude of PHE increases with field strength following a $B^2$ law because of the quadratic $\rho_{zz}(B)$ as well as $\Delta\rho(B)$ \cite{Nandys-PRL2017}. Such a rule seems obeyed in CeCo$_2$Ga$_8$ only for low field $B\leq 5$ T. High-field saturation of PHE has been observed in Dirac semimetal ZrTe$_{5-\delta}$ and discussed in the framework of chiral anomaly with $L_a<L_x<L_c^2/L_a$ where $L_x$, $L_a$ and $L_c$ are respectively sample length, ``chiral" magnetic length ($L_a\propto 1/B$) and chiral charge diffusion length \cite{P.Li ZrTe,BurkovAA-PRB2017}. However, this mechanism is unlikely in CeCo$_2$Ga$_8$ due to the irrelevance of band topology. Regardless of the unclear mechanism to quantitatively understand the high-field saturation of PHE, we notice another salient feature, i.e., the maximum of $\rho_{zz}(B)$ and the deviation of PHE from the $B^2$-law seem to be synchronous. This reminds us that the two may have the same origin - a competition between incoherent and coherent Kondo scatterings. To see this more clearly, temperature dependence of PHE investigation will be helpful.

Figure 3 displays PHE of CeCo$_2$Ga$_8$ at different temperatures, under a fixed field 15 T. We start with the results at 80 K, just below $T_K^{on}$. A two-fold oscillation pattern, albeit weakly distorted from a perfect sinusoidal shape, is clearly seen at this temperature, cf. the left panel of Fig.~3(a). A polar plot is also presented in the right panel. It should be mentioned that the plume orients along $\varphi\approx45~^{\circ}$, which happens to be the angle where $\rho_{xz}^\text{PHE}$ minimizes at 1.6 K. This implies that a sign reversal (or equivalently $180 ^{\circ}$ phase shift) may have taken place across $T^*$. In between, the evolution of PHE with decreasing $T$ is more complicated than expected. Additional small peaks show up in $\rho_{xz}^\text{PHE}(\varphi)$, leading to the severely distorted PHE pattern which is most apparent near 20 K in Fig.~3(d). To see this feature more clearly, we performed Fast Fourier Transform (FFT) on $\rho_{xz}^{\text{PHE}}(\varphi)$ [Fig.~3(i)]. Besides the main oscillatory component $A_2$, a series of higher-order components $A_n$ (higher harmonics with $n=4,6...$) are also visible. Further cooled down, a new peak positioned near 135 $^\circ$ grows rapidly in $\rho_{xz}^{\text{PHE}}(\varphi)$, while the original peak shrinks and is gradually taken over by the new peak. At 5 K (below $T^*$), $\rho_{xz}^\text{PHE}(\varphi)$ restores two-fold oscillations, whereas the principal axis of the plume is shifted by about $90~^\circ$ as compared with that of 80 K, seeing Fig.~3(h). The entire process can also be reflected in the temperature dependence of $A_n$ as shown in Fig.~3(j), where one can observe a pronounced drop of $A_2$ between $T_K^{on}$ and $T^*$, whereas the relative weights of the higher-order $A_n$s rise up below $T_K^{on}$ and peak just above $T^*$ (cf. $A_4/|A_2|$ and $A_6/|A_2|$). As the temperature further decreases from $T^*$, these higher harmonics also diminish, while $A_2$ crosses zero and then varies monotonically to attain a large negative value.

Note that although the sign of PHE reverses between incoherent and coherent regimes, $\Delta\rho$ ($=\rho_{zz}^{\mathbf{B}\parallel[100]}-\rho_{zz}^{\mathbf{B}\parallel[001]}$) remains positive across the full window 1.6-50 K, as shown in Fig.~4(a-c). This provides additional evidence for the nonlinearity between the amplitudes of PHE and PAMR. Combining Figs.~2 and 4, a straightforward conclusion can be drawn, i.e. the sign change of PHE here is not simply related to the sign change of $\Delta\rho$; in other words, the PHE observed in CeCo$_2$Ga$_8$ seems far beyond the classic description given by Eqs.~(1). \\

\textbf{DISCUSSION}

To summarize the main experimental observations here, the abnormal PHE of CeCo$_2$Ga$_8$ is featured by: (\rmnum{1}) rapid decrease of two-fold oscillation and increase of higher harmonics for $T$ between $T_K^{on}$ and $T^*$, (\rmnum{2}) suppression of higher harmonics and emergence of a new two-fold oscillation mode of opposite sign below $T^*$, and (\rmnum{3}) loss of linearity between the oscillating amplitudes of PHE and PAMR. Natural questions, then, concern the origin of this abnormal PHE below $T_K^{on}$ and the emergent new mode below $T^*$.

Abnormal PHEs with either distorted shapes or additional oscillations have been reported in many systems, e.g. Bi$_4$Br$_4$ \cite{ZhongJ-Bi4Br4ACSNano2024}, TaP \cite{YangJ-PRM2019}, LaAlO$_3$/SrTiO$_3$ (111) interface \cite{RoutPK-PRB2017}, bismuth \cite{YangSY-PRR2020}, TbPtBi \cite{ZhuY-PRB2020}, MnBi$_2$Te$_4$ \cite{WuM-NanoLett2022}, KV$_3$Sb$_5$ \cite{LiL-npjQM2023}, etc. In general, the mechanisms for abnormal PHE can be categorized as follows: (\rmnum{1}) an extrinsic factor caused by current jetting, which is usually seen in high-mobility semimetals such as TaP \cite{YangJ-PRM2019}; (\rmnum{2}) phase coherence effect in topologically nontrivial materials with weak antilocalization, as exemplified by Bi$_4$Br$_4$ \cite{ZhongJ-Bi4Br4ACSNano2024}; (\rmnum{3}) orbital anisotropy like in bismuth \cite{YangSY-PRR2020}; (\rmnum{4}) competition in transport between bulk and surface states mediated by orbital magnetic moment in the magnetic topological insulator MnBi$_2$Te$_4$ \cite{WuM-NanoLett2022}; (\rmnum{5}) coupling between fluctuations from lattice, spin, electronic and other degrees of freedom, seeing KV$_3$Sb$_5$ for instance \cite{LiL-npjQM2023}.

In the case of CeCo$_2$Ga$_8$, firstly, no topological nature was reported in first-principles calculations \cite{WangL-CeCo2Ga8}, and our previous transport measurements did not find any signature of high mobility in this compound either \cite{ChengK-CeCo2Ga8Q1D,ChengK-CeCo2Ga8Stress}. Therefore, interpretations related to current-jetting or band topology are not applicable. Secondly, the loss of linearity between PHE amplitude and $\Delta\rho$ implies that there are multiple scattering mechanisms in transport in CeCo$_2$Ga$_8$, and they contribute differently to magnetoresistance and PHE. The commonest mechanism for magnetoresistance arises from the bending of carrier trajectory under Lorentz force, which typically results in Kohler's rule in conventional metals, viz MR measured at different fields and temperatures can be scaled into a single function of $\rho_{zz}(0)/B$. However, such an attempt turns out to be a failure for both $\mathbf{B}\parallel[100]$ and $[001]$, seeing Figs.~4(e,f). This manifests that regular scattering process for orbital magnetoresistance solely can hardly explain the results in CeCo$_2$Ga$_8$. On the other hand, the coincidence of the development and suppression of the higher-order oscillation patterns in PHE with the onset of incoherent Kondo scattering at $T_K^{on}$ and the coherence temperature $T^*$ in the resistivity strongly suggests key roles played by the incoherent and coherent Kondo scatterings. For this purpose, the scaling behavior based on the Bethe-ansatz solution of the Coqblin-Schrieffer model was tested \cite{CoqblinB-PR1969,SchlottmannP-PhysRep1989},
\begin{equation}
\mathrm{MR}=f(\frac{B}{T+T_K}),
\label{Eq3}
\end{equation}
where $T_K$ is the single-ion Kondo temperature. This model is found to fit the MR$(T,B)$ curve rather well in the incoherent Kondo regime in that all the curves collapse onto a common line [Figs.~4(g,h)], yielding $T_K=19(5)$ K. Note that the gain of magnetic entropy recovers $R\ln{2}$ near 19 K \cite{WangL-CeCo2Ga8}, further validating our Coqblin-Schrieffer scaling analysis. Please be also noted that this scaling law remain valid until well below $T^*$, indicating that incoherent Kondo scattering probably still plays some role even after the onset of Kondo coherence. For sufficiently low $T$, coherent Kondo scattering becomes dominant as evidenced by the quasi-$B^2$ law of MR [inset to Fig.~2(b)]. Correspondingly, the higher harmonics ($A_n$ for $n\ge 4$) are suppressed and the new $A_2$ mode of negative sign rises up as shown in Fig.~3(j). These coincidences motivate us to associate the abnormal PHE with the incoherent Kondo scattering between $T_K^{on}$ and $T^*$, and the new mode with the coherent Kondo scattering below $T^*$. Their coexistence below $T^*$ reflects the coexistence of residual unscreened local moments with incoherent Kondo scattering and the coherent heavy-electron state predicted by the two-fluid theory \cite{Nakatsuji-PRL2004,Yang-PRL2008}, whose relative weights are temperature dependent and can be tuned by external parameters like the magnetic field.

A schematic cartoon is proposed in Fig.~5 based on this idea. The phase diagram is divided into three parts: Localized regime, Incoherent Kondo regime (or Dynamic hybridization regime), and Coherent Kondo regime. At high temperature above $T_K^{on}$, the $f$ electrons are fully localized, behaving as individual magnetic scattering centers, and the FS is totally determined by the conduction bands which in CeCo$_2$Ga$_8$ are some cylindrical sheets along $\Gamma$-Y \cite{WangL-CeCo2Ga8}. The PHE then exhibits the usual two-fold oscillatory pattern from conduction electrons as predicted by the conventional theory for a normal metal. As the temperature is decreased to $T_K^{on}$, incoherent Kondo scattering starts to play a role as indicated by the logarithmic temperature dependence in the resistivity in Fig.~1(c). The PHE then displays abnormal patterns due to higher harmonics induced by incoherent Kondo scattering from the local $f$-moments. The complex angle-dependence might be associated with anisotropic magnetic scattering due to strong spin-orbit coupling of the $f$-electron crystalline ground state \cite{RoutPK-PRB2017,ZhuY-PRB2020,ChengK-CeCo2Ga8Q1D}. With lowering temperature, the relative weights of these higher harmonics increase while the overall magnitude of the two-fold oscillation $A_2$ decreases, implying increasingly strong influences of the incoherent Kondo scattering on conduction electrons. This explains why ARPES experiments have observed gradual bending of the conduction bands while ultrafast optical spectroscopy measurements revealed precursor hybridization fluctuations in other heavy-fermion metals between $T_K^{on}$ and $T^*$ \cite{ChenQY-PRB2017,KoitzschA-PRB2008,LiuY-PRL2020,JangaS-CeCoIn5PNAS2020}. We propose to perform similar experiments for CeCo$_2$Ga$_8$. As the temperature is further lowered to about $T^*$, coherence sets in, causing rapid suppression of the abnormal PHE. Correspondingly, a new two-fold oscillation mode arises due to the emergent coherent heavy-electron state whose weight grows rapidly with lowering temperature. At sufficiently low temperatures, where the coherent heavy-electron state takes over and the Fermi surfaces are enlarged to incorporate $f$-electrons, the PHE recovers the two-fold oscillation pattern, albeit with a phase shift determined by the anisotropy of the enlarged Fermi surfaces. Indeed, parts of the Fermi surfaces change topology from cylindrical sheets of the conduction bands to flat sheets of heavy electrons due to the quasi-1D Kondo chain structure of CeCo$_2$Ga$_8$ \cite{WangL-CeCo2Ga8}. This scenario also explains why the most complicated pattern appears near 20 K (at 15 T) where coherence already sets in for $\mathbf{B}\parallel[001]$ but not for $\mathbf{B}\parallel[100]$, cf. Fig.~4(d).

For a better intuitive understanding, we performed a semi-quantitative simulation based on the above hypothesis. As predicted by the two-fluid theory, we may write down the total PHE conductivity to arise from two conducting channels, namely the conduction electron contribution from incoherent Kondo scattering and the coherent heavy electron contribution. This yields (more details are provided in \textbf{SI})
\begin{equation}
\begin{aligned}
\sigma_{zx}^{\text{PHE}}=\left\{
  \begin{array}{ll}
    \Delta\sigma^c \sin{\varphi}\cos{\varphi}  &~~(T>T_K^{on}) \\
    \Delta\sigma^c\left(\sin{\varphi}\cos{\varphi}+a_4\sin{2\varphi}\cos{2\varphi}\right)                &~~(T^*<T<T_K^{on})                \\
    (1-w)    \Delta\sigma^c\left(\sin{\varphi}\cos{\varphi}+a_4\sin{2\varphi}\cos{2\varphi}\right)  +w    \Delta\sigma^h\sin{\varphi}\cos{\varphi}       &~~(T<T^*)     \\
   \end{array}
\right.
\end{aligned}
\end{equation}
where $\Delta\sigma^{c,h}$ are the differences of the electrical conductivity parallel with and perpendicular to the magnetic field \cite{ZhongJ-CPB2023}, and the superscripts $c$ and $h$ represent the contributions from the conduction bands ($c$) and the heavy-electron quasiparticles ($h$), respectively. A temperature dependent weight $w(T)$ is employed to characterize the evolution of the two conducting channels. In accordance with our PAMR results, $\Delta\sigma^c>0$, $\Delta\sigma^h<0$, and $\sigma^h\ll\sigma^c$ are presumed. For simplicity, we only consider a single higher-order oscillation mode $A_4$ from the incoherent Kondo scattering ($a_4\equiv A_4/A_2$). Figure 5 presents some typical numerical results, where the calculated $\sigma_{zx}^{\text{PHE}}$ has been converted back to PHE resistivity $\rho_{xz}^{\text{PHE}}$. One can clearly observe the sign reversal of $\rho_{xz}^{\text{PHE}}$ for temperatures above $T_{K}^{on}$ and below $T^*$. Our simplest numerical simulation indeed captures all the major features of PHE in CeCo$_2$Ga$_8$.

It should be pointed out that although resistive anomaly (e.g. $\rho \propto -\ln T$) has been reported to characterize the incoherent Kondo scattering, the PHE reveals for the first time such peculiar oscillation patterns for Kondo lattice systems, which allows us to further disentangle the contributions from incoherent and coherent scatterings below $T^*$ and provide clear transport evidence for the two-fluid theory. By contrast, a regular Hall effect (out-of-plane $\mathbf{B}\parallel\mathbf{j}$) is nearly featureless in this regime, seeing Fig.~S4 in \textbf{SI}. More related studies are encouraged in other prototypical heavy-fermion systems such as Ce115s and so on. Theoretically, quantitative descriptions are also required to interpret the microscopic nature of the electronic structure and scattering mechanism in the whole temperature range.

In all, longitudinal and transverse resistivities of the quasi-1D Kondo lattice compound CeCo$_2$Ga$_8$ are systematically investigated in the presence of rotational in-plane field, and abnormal PHE and PAMR that substantially deviate from conventional empirical formulas are observed. Of particular interest is the temperature evolution of the higher harmonics and the  emergent two-fold oscillatory mode at low temperatures. By utilizing their different oscillation patterns, our work provides an effective way to disentangle the contributions of incoherent Kondo scattering and coherent heavy-electron state, and to track the microscopic evolution of the Kondo lattice coherence including a possible precursor state with dynamic hybridization. These findings also imply that PHE - which used to be mostly considered for potential application in magnetic sensors - can be also informative for clarifying the nature of localized-itinerant transition in strongly-correlated quantum materials.\\

\textbf{METHODS}\\

\textbf{Single crystals preparation and characterization}

Single crystalline CeCo$_{2}$Ga$_8$ was grown in a two-step method \cite{WangL-CeCo2Ga8}. First, polycrystalline CeCo$_2$Ga$_8$ was prepared by arc-melting a stoichiometric mixture of Ce, Co and Ga. Second, Gallium self-flux method was used to grow CeCo$_{2}$Ga$_8$ single crystals. This was carried out by mixing up CeCo$_{2}$Ga$_8$ polycrystal and Ga in an alumina crucible in a molar ratio of 1:10, and sealed in an evacuated quartz tube. The latter was heated up to 1100 $^{\circ}$C in 10 h, dwelt for 10 h, and slowly cooled at a rate of 2 $^{\circ}$C/h down to 630 $^{\circ}$C when the excess flux was removed by centrifugation. Milimeter-sized CeCo$_2$Ga$_8$ single crystals can be obtained. The chemical composition, crystalline structure and orientation of the selected crystal were verified by energy dispersive spectroscopy (EDS), Wavelength-dispersive X-ray (WDX) spectroscopy, X-ray diffraction (XRD) and Laue XRD, seeing \textbf{SI}. \\

\textbf{Device fabrication and transport measurements}

The microstructured Hall-bar device was fabricated by FIB technique (Scios 2 HiVac, Thermo Fisher Scientific Inc.) with effective dimensions $50\times15\times8$ $\mu$m$^3$ along the principal axes [001], [100] and [010] respectively. Electrical contacts were made via patterned mask with photolithography subsequent growth of Ti/Au (10 nm / 100 nm) layers. Electrical transport experiments were performed in an IntegraAC Mk \Rmnum{2} recondensing cryostat equipped with a 16 T superconducting magnet (Oxford Instruments). The electrical current was applied along [001]. Both in-plane longitudinal resistivity \lyk{[$\rho_{zz}~(\equiv E_z/j_z)$]} and transverse resistivity \lyk{[$\rho_{xz}~(\equiv E_x/j_z)$]} were measured\lyk{, where $E_{z,x}$ are the specific components of the electric field that were obtained by directly measuring the voltages $V_{z,x}$ with} an AC Resistance bridge (372, Lake Shore Cryotronics). A home-built rotation probe was exploited to conduct the in-plane 0-360 $^{\circ}$ field rotation with respect to the current. \\

\textbf{Data availability}

The data supporting the findings of this study are available from the corresponding authors upon reasonable request via email to Y. Luo. \\

\emph{}\\
\textbf{ACKNOWLEDGEMENTS}\\

The authors thank Joe D. Thompson for helpful discussions. This work is supported by the National Key R\&D Program of China (2022YFA1602602, 2022YFA1402203, 2023YFA1609600), National Natural Science Foundation of China (U23A20580, U22A6005, U2032204, 12174429), the open research fund of Songshan Lake Materials Laboratory (2022SLABFN27), Guangdong Basic and Applied Basic Research Foundation (2022B1515120020), Beijing National Laboratory for Condensed Matter Physics (2024BNLCMPKF004), and the Synergetic Extreme Condition User Facility (SECUF).
\\

\textbf{AUTHOR CONTRIBUTIONS}

Y.L. conceived and designed the experiments. Y.S. and G.D. grew and characterized the crystals. X.G. and Z.Z. fabricated the micro-structured device with FIB. S.Z. performed most of the measurements with the aids from H.Z., Z.W. and F.L. S.Z., Y.Y. and Y.L. discussed the data, interpreted the results, and wrote the paper with input from all the authors. \\

\textbf{COMPETING INTERESTS}

The authors declare no competing interests.\\

\textbf{ADDITIONAL INFORMATION}\\

\textbf{Reprints and permission information} is available at \\
http:$\backslash\backslash$www.nature.com/reprints \\

\textbf{Publisher's note} Springer Nature remains neutral with regard to jurisdictional claims in published maps and institutional affiliations.

\newpage
\begin{figure}[!htp]
\hspace*{-0pt}
\vspace*{-0pt}
\includegraphics[width=16.5cm]{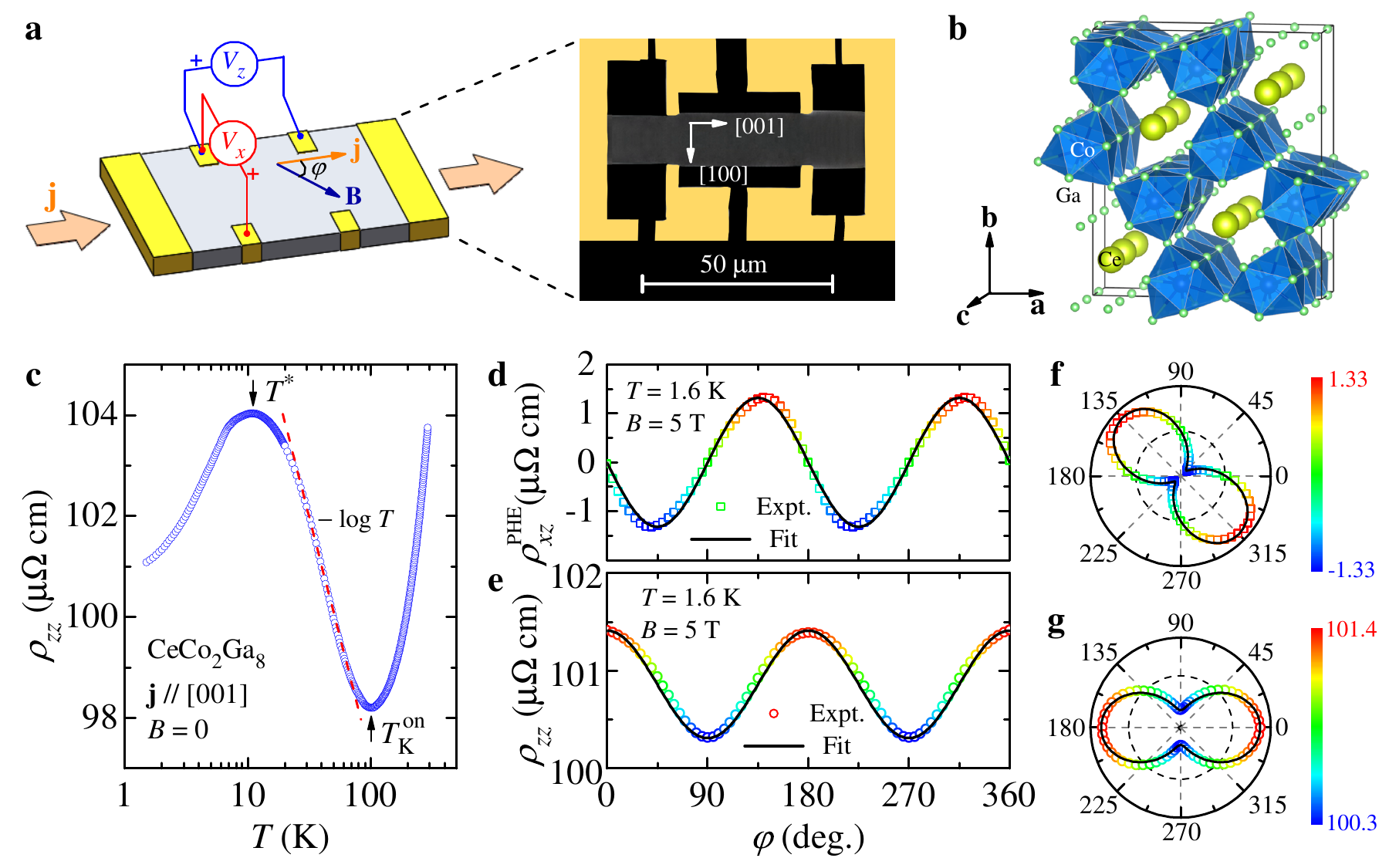}
\label{Fig1}
\end{figure}
\vspace*{-20pt}
\textbf{Fig. 1 $|$ Planar Hall Effect (PHE) and Planar Anisotropic Magnetoresistance (PAMR) of CeCo$_2$Ga$_8$.} \textbf{a} Sketch of PHE [$\rho_{xz}^\text{PHE}(\varphi)$] and PAMR [$\rho_{zz}(\varphi)$] measurements. $\varphi$ characterizes the angle spanned by electrical current $\mathbf{j}$ and external magnetic field $\mathbf{B}$. The sample investigated is prepared by focused ion beam (FIB) with $\mathbf{j} \parallel$ [001], and the rotation of $\mathbf{B}$ is within $\mathbf{ac}$ plane. \textbf{b} Crystalline structure of CeCo$_2$Ga$_8$ which shows quasi-1D cerium chain along [001]. \textbf{c} Temperature dependence of resistivity $\rho_{zz}$. The onset of Kondo effect is denoted by $T_K^{on}\approx100$ K, and the Kondo coherence develops below $T^*\approx12$ K where $\rho_{zz}$ peaks. The red dashed line is a fit to the $-\log T$ law in the incoherent regime. \textbf{d} $\rho_{xz}^\text{PHE}(\varphi)$ at 1.6 K. The symbols represent experimental data, while the solid line is the fitting curve. \textbf{e} $\rho_{zz}(\varphi)$ at 1.6 K. \textbf{f} and \textbf{g} respectively show $\rho_{xz}^\text{PHE}(\varphi)$ and $\rho_{zz}(\varphi)$ in polar coordinates that exhibit two-fold oscillations. The color bars signify the values of $\rho_{xz}^\text{PHE}(\varphi)$ and $\rho_{zz}(\varphi)$.  \\

\newpage
\begin{figure*}[!htp]
\vspace*{0pt}
\hspace*{0pt}
\includegraphics[width=14cm]{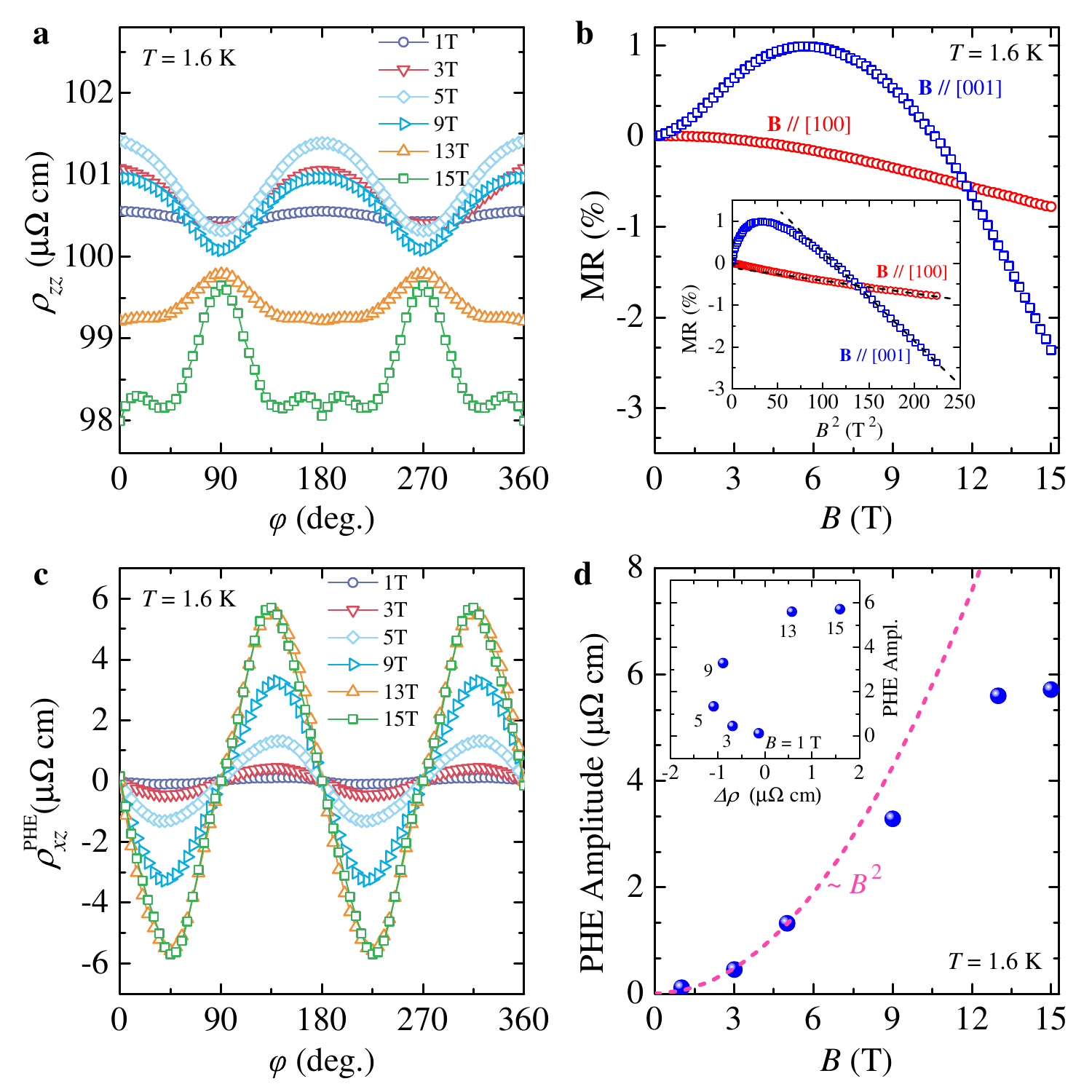}
\label{Fig2}
\end{figure*}
\vspace*{-0pt}
\textbf{Fig. 2 $|$ Planar Anisotropic Magnetoresistance (PAMR) and Planar Hall Effect (PHE) of CeCo$_2$Ga$_8$ under various magnetic fields.} \textbf{a} $\rho_{zz}(\varphi)$ for different field strengths. The two-fold oscillation persists until field larger than 9 T. \textbf{b} Magnetoresistance as a function of field for $\mathbf{B}\parallel[100]$ (red circles) and $\mathbf{B}\parallel[001]$ (blue squares). The inset shows MR in the $B^2$ scale. \textbf{c} $\rho_{xz}^\text{PHE}(\varphi)$ curve at 1.6 K for different field strength. All the curves exhibit two-fold oscillation. \textbf{d} PHE amplitude as a function of magnetic field. A quadratic-$B$ law fits the curve rather well for field less than 5 T, while for larger field, the amplitude of PHE tends to saturate. The inset is a plot of PHE amplitude vs. $\Delta\rho$ where no systematic linear relationship can be found between them.\\

\newpage
\begin{figure*}[!htp]
\vspace*{-0pt}
\hspace*{-0pt}
\includegraphics[width=16.5cm]{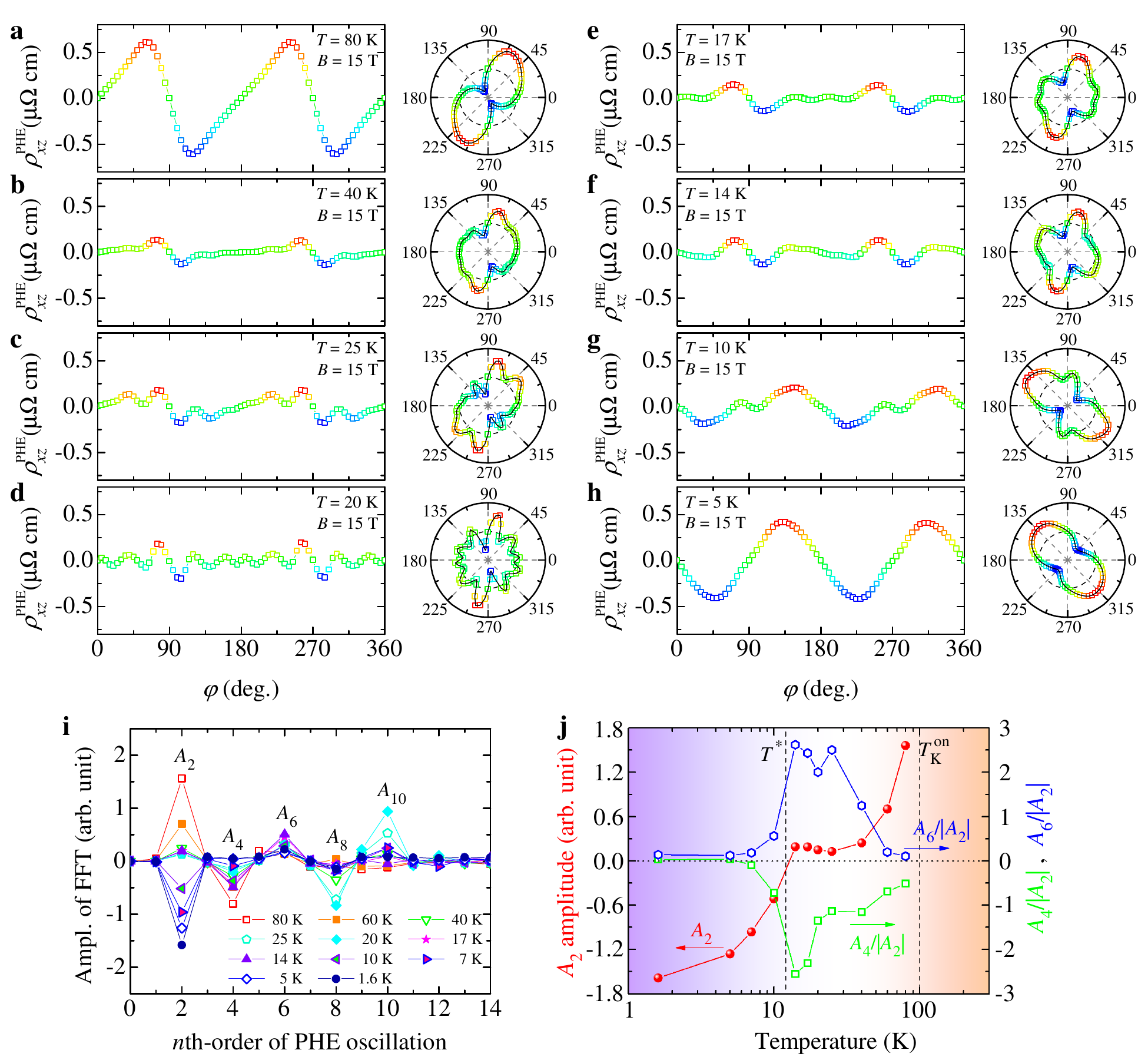}
\label{Fig3}
\end{figure*}
\vspace*{-20pt}
\textbf{Fig. 3 $|$ Planar Hall Effect (PHE) of CeCo$_2$Ga$_8$ at various temperatures.} Measurements were made under fixed magnetic field $B=15$ T. $\rho_{xz}^\text{PHE}(\varphi)$ profiles at representative temperatures below $T_K^{on}$: \textbf{a} 80 K; \textbf{b} 40 K; \textbf{c} 25 K; \textbf{d} 20 K; \textbf{e} 17 K; \textbf{f} 14 K; \textbf{g} 10 K; \textbf{h} 5 K. The plume orients along $\varphi\approx45^{\circ}$ at 80 K. As $T$ decreases, additional oscillations appear, and when $T$ is well below $T^*$, the two-fold oscillations restore with the principal axis along $\varphi\approx135^{\circ}$. \textbf{i} Fast Fourier Transform (FFT) spectrum of $\rho_{xz}^{\text{PHE}}(\varphi)$ which gives rise to $A_{n}$, the amplitude of the $n$-th order of the PHE oscillation. \textbf{j} Left, temperature dependence of $A_2$; right, $A_4/|A_2|$ and $A_6/|A_2|$ as functions of $T$. \\

\newpage
\begin{figure*}[!htp]
\vspace*{-0pt}
\hspace*{-0pt}
\includegraphics[width=14cm]{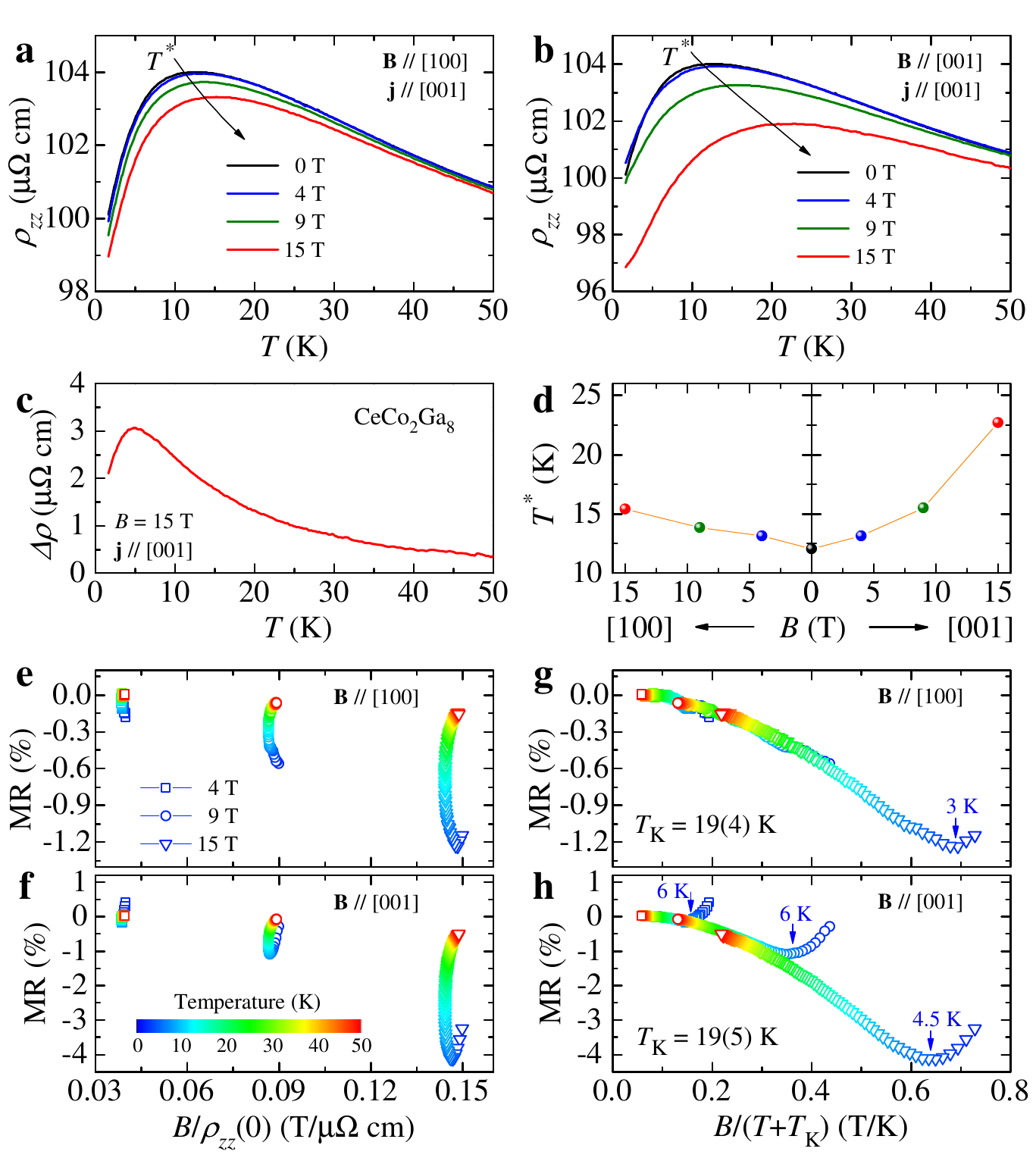}
\label{Fig4}
\end{figure*}
\vspace*{-0pt}
\textbf{Fig. 4 $|$ Magnetoresistance (MR) of CeCo$_2$Ga$_8$.} \textbf{a} Temperature dependence of $\rho_{zz}$ under various fields 0, 4, 9 and 15 T, $\mathbf{B}\parallel[100]$. $\textbf{b}$ Temperature dependence of $\rho_{zz}$ under various fields $\mathbf{B}\parallel[001]$. \textbf{c} $\Delta\rho=\rho_{zz}^{\mathbf{B}\parallel[100]}-\rho_{zz}^{\mathbf{B}\parallel[001]}$ as a function of $T$, under external field 15 T. \textbf{d} A plot of $T^*$ vs. $B$ for both field directions. \textbf{e,f} Kohler's plot. MR as a function of $B/\rho_{zz}(0)$ for $\mathbf{B}\parallel[100]$ and $\mathbf{B}\parallel[001]$, respectively. For both field directions, Kohler's rule is drastically violated. \textbf{g,h} Coqblin-Schrieffer analysis for $S=1/2$ Kondo impurity model. MR vs. $B/(T+T_K)$. For both field directions, MR data in the hybridization dynamic regime fall on a single line. Note that such a scaling law holds even for below $T^*$.  \\

\newpage
\begin{figure*}[!htp]
\vspace*{-0pt}
\hspace*{-0pt}
\includegraphics[width=16.5cm]{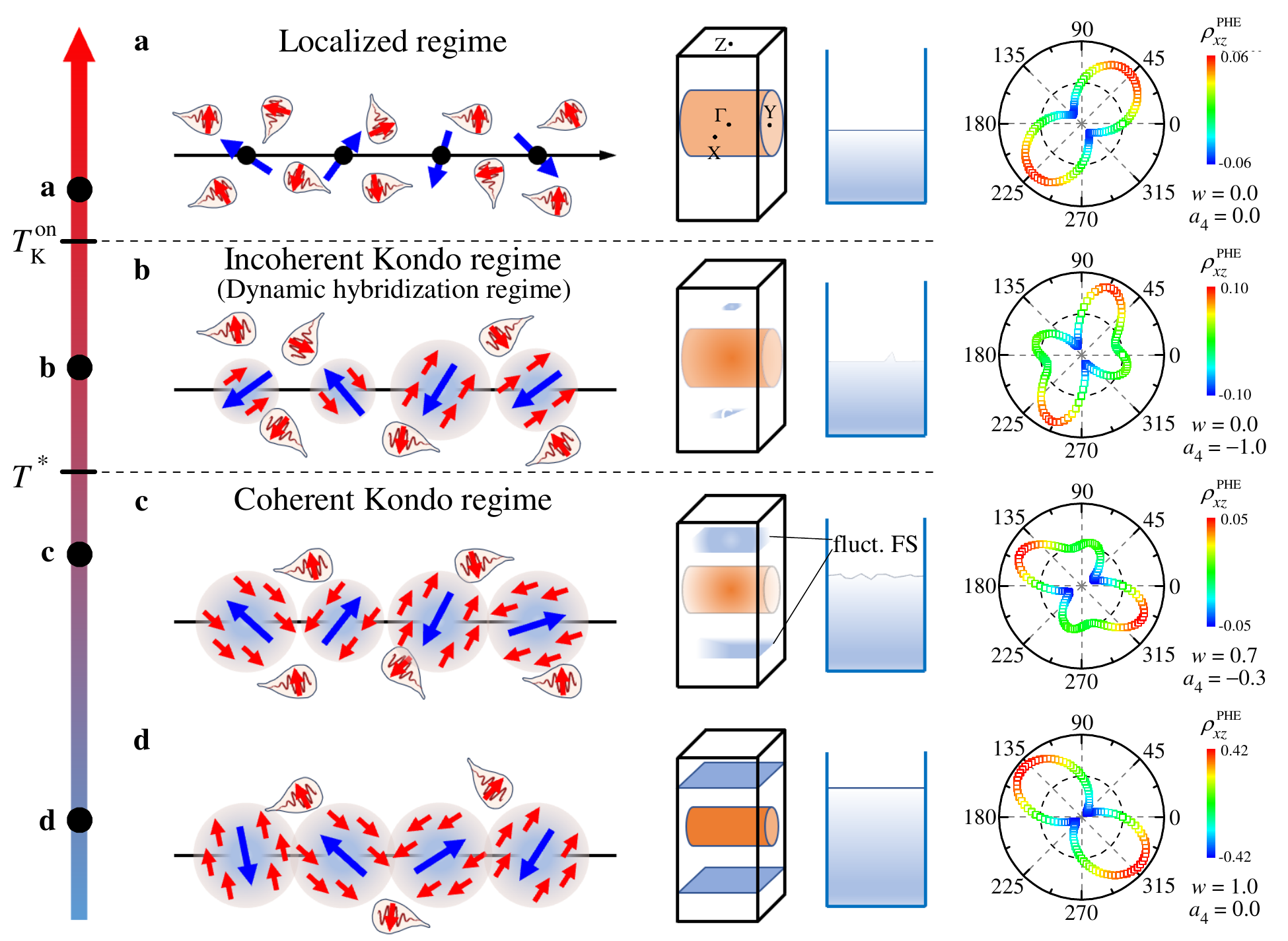}
\label{Fig5}
\end{figure*}
\vspace*{-30pt}
\textbf{Fig. 5 $|$ Cartoon illustration about the establishment of heavy-electron state and simulations of Planar Hall Effect (PHE).} The schematic Fermi surface is from theoretical calculations in Ref.~\cite{WangL-CeCo2Ga8}. The hatched balls signify Kondo entanglement of $f$ (blue) and conduction (red) electrons. Above incoherent Kondo scale $T_K^{on}$, the $f$ electrons are fully localized, and only light-mass cylindrical sheets from Co/Ga orbitals are present along $\Gamma$-Y. For $T^*<T<T_K^{on}$, the $c$-$f$ hybridization is incoherent. Below coherent Kondo scale $T^*$, Kondo coherence sets in, two fluids coexist, and fluctuating new flat sheets with heavy mass start to appear. Well below $T^*$, coherent heavy-electron state is stabilized. The right-column panels are simulations based on this hypothesis. The simulations are made with the parameters $\Delta\sigma^c=0.10$, and $\Delta\sigma^h=-0.04$. Temperature dependent $w(T)$ and $a_4$ are employed to characterize the weight of transport conductivity contributed by the heavy fermion quasiparticles and the four-fold oscillation mode: \textbf{a} $w=0.0$ and $a_4=0.0$ for $T>T_K^{on}$; \textbf{b} $w=0.0$ and $a_4=-1.0$ for $T^*<T<T_K^{on}$; \textbf{c} $w=0.7$ and $a_4=-0.3$ for $T<T^*$; and \textbf{d} $w=1.0$ and $a_4=0.0$ for well below $T^*$. The color bars signify the values of $\rho_{xz}^\text{PHE}(\varphi)$.\\

\clearpage

\renewcommand{\thefigure}{S\arabic{figure}}
\renewcommand{\thetable}{S\arabic{table}}
\renewcommand{\theequation}{S\arabic{equation}}
\setcounter{table}{0}
\setcounter{figure}{0}
\setcounter{equation}{0}
\setcounter{page}{1}

\vspace{-15pt}

\begin{center}
\large
\textbf{Supplementary Information: } \\
\textbf{Abnormal planar Hall effect and disentanglement of incoherent and coherent transport in a Kondo lattice}\\
\small
\emph{}\\
Shuo Zou$^{1}$, Hai Zeng$^{1}$, Zhuo Wang$^{1}$, Guohao Dong$^{2}$, Xiaodong Guo$^{1}$, Fangjun Lu$^{1}$, Zengwei Zhu$^{1}$,  Youguo Shi$^{2,3*}$, Yi-feng Yang$^{2,3\dag}$, and Yongkang Luo$^{1\ddag}$ \\
$^1$ {\it Wuhan National High Magnetic Field Center and School of Physics, Huazhong University of Science and Technology, Wuhan 430074, China;}\\
$^2$ {\it Beijing National Laboratory for Condensed Matter Physics, Institute of Physics, Chinese Academy of Sciences, Beijing 100190, China; and}\\
$^3$ {\it University of Chinese Academy of Sciences, Beijing 100049, China}\\
\end{center}

\normalsize

In this \textbf{Supplementray Information (SI)}, we provide additional results that will further support the discussions and conclusion in the main text, including sample characterization, PHE results at more temperatures, numerical simulations of PHE, Hall effect etc.\\

\textbf{SI \Rmnum{1}: Sample characterization}\\

\begin{figure*}[!htp]
	\vspace*{-34pt}
	\includegraphics[width=15.3cm]{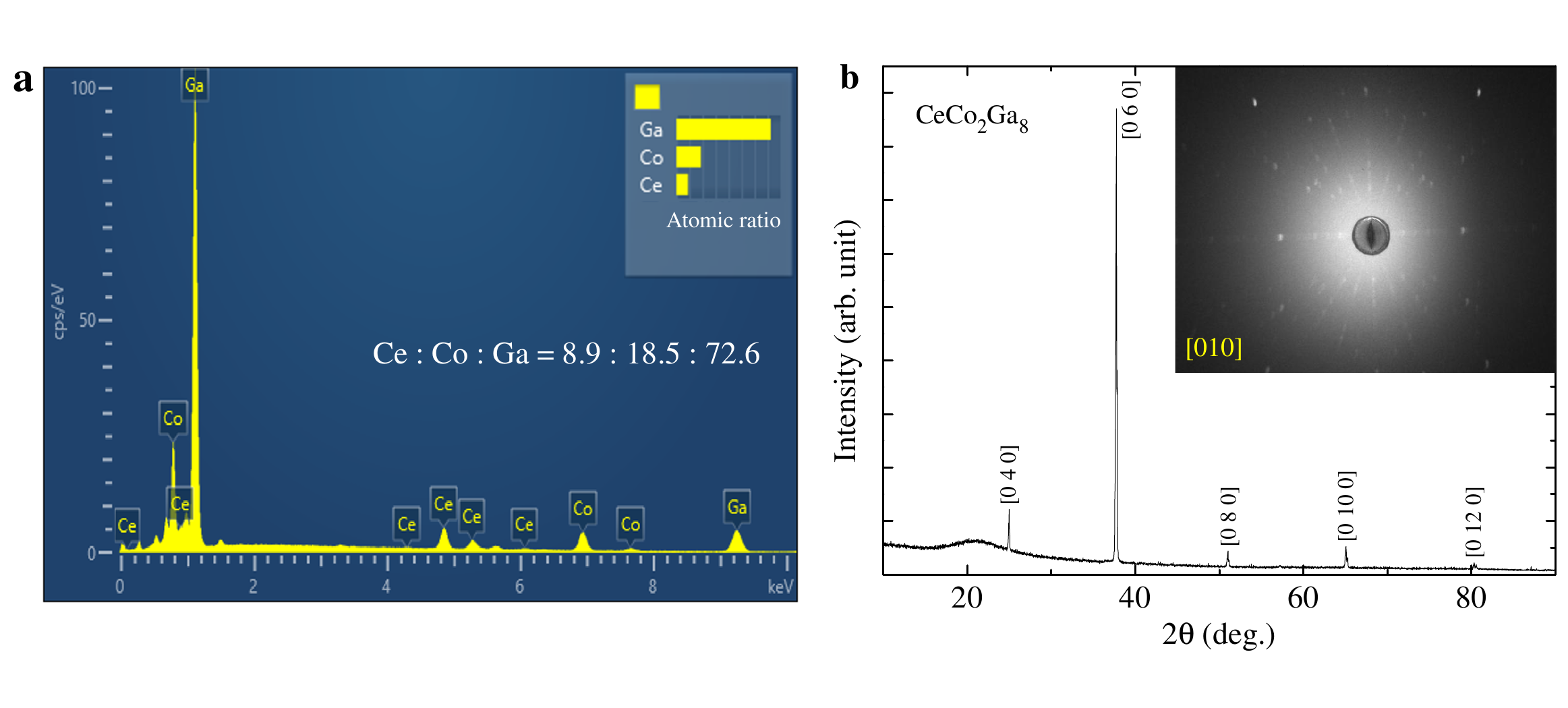}
	\label{Fig.S1}
\vspace*{-20pt}
\small
\begin{flushleft}
\justifying{
    \textbf{Fig. S1 $|$ Sample characterization of CeCo$_2$Ga$_8$.} \textbf{a} EDS results, which indicate the atomic ratio Ce : Co : Ga $\approx$ 8.9 : 18.5 : 72.6. \textbf{b} Room-temperature XRD pattern on the polished $\mathbf{ac}$ plane, where only $[0, 2l, 0]$ peaks are observed. The inset shows Laue XRD results. The nice pattern on the [010] surface guarantees the good crystallization. No signature of multi-grain or impurity can be seen. The extent of misalignment is estimated to be less than $\sim 1^{\circ}$.
    }
\end{flushleft}
\normalsize
\end{figure*}

\begin{figure*}[!htp]
	\vspace*{-10pt}
	\includegraphics[width=15cm]{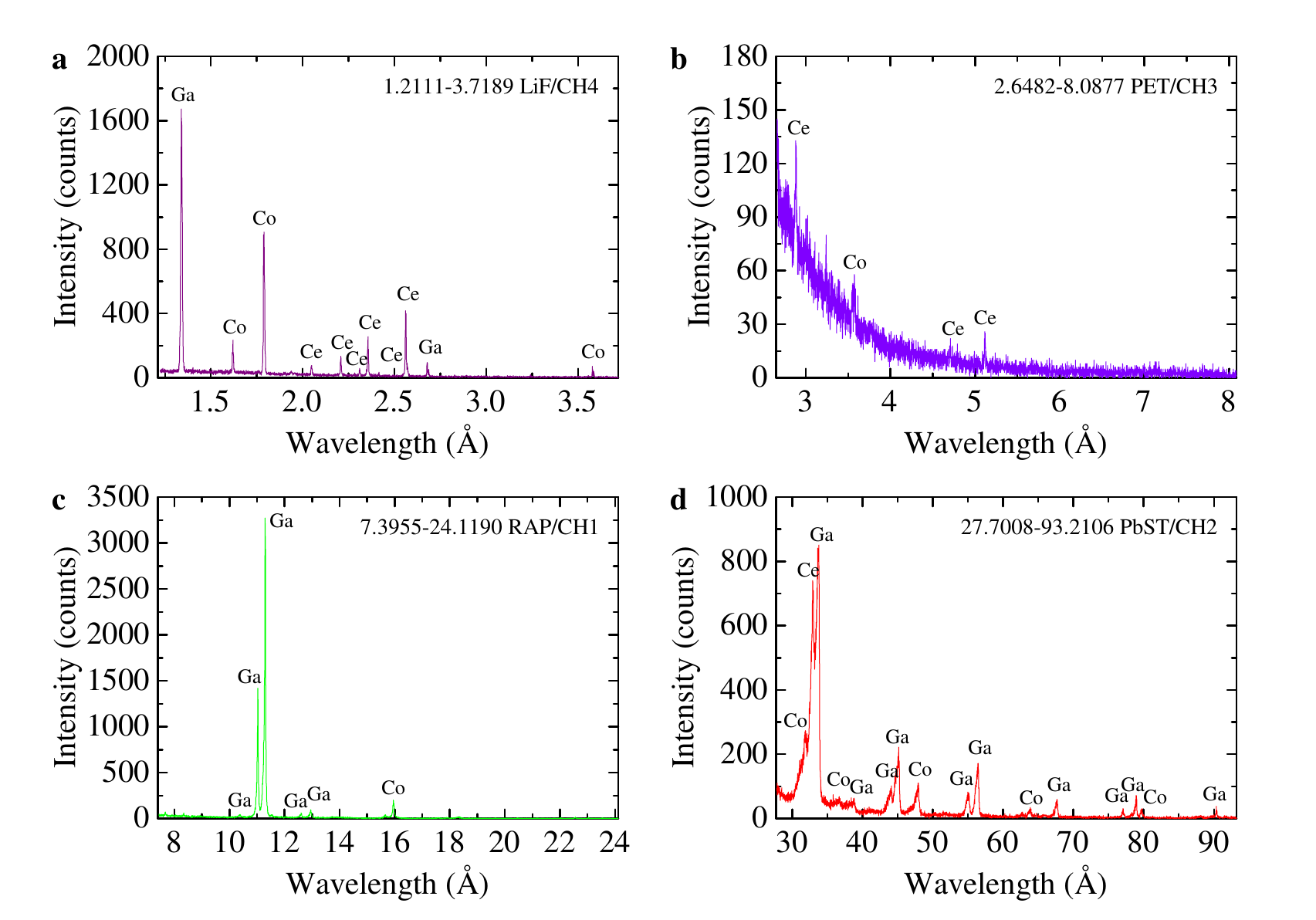}
	\label{Fig.S2}
\vspace*{-10pt}
\small
\begin{flushleft}
\justifying{
    \textbf{Fig. S2 $|$ Wavelength-dispersive X-ray (WDX) spectra of CeCo$_2$Ga$_8$.} The absence of any discernible extraneous peaks demonstrates the high purity of the sample. A quantitative analysis leads to the molar ratio Ce : Co : Ga $\approx$ 9.18 : 19.34 : 71.48, close to the stoichiometry of CeCo$_2$Ga$_8$.
    }
\end{flushleft}
\normalsize
\end{figure*}

\begin{figure*}[!htp]
	\vspace*{-0pt}
	\includegraphics[width=16.5cm]{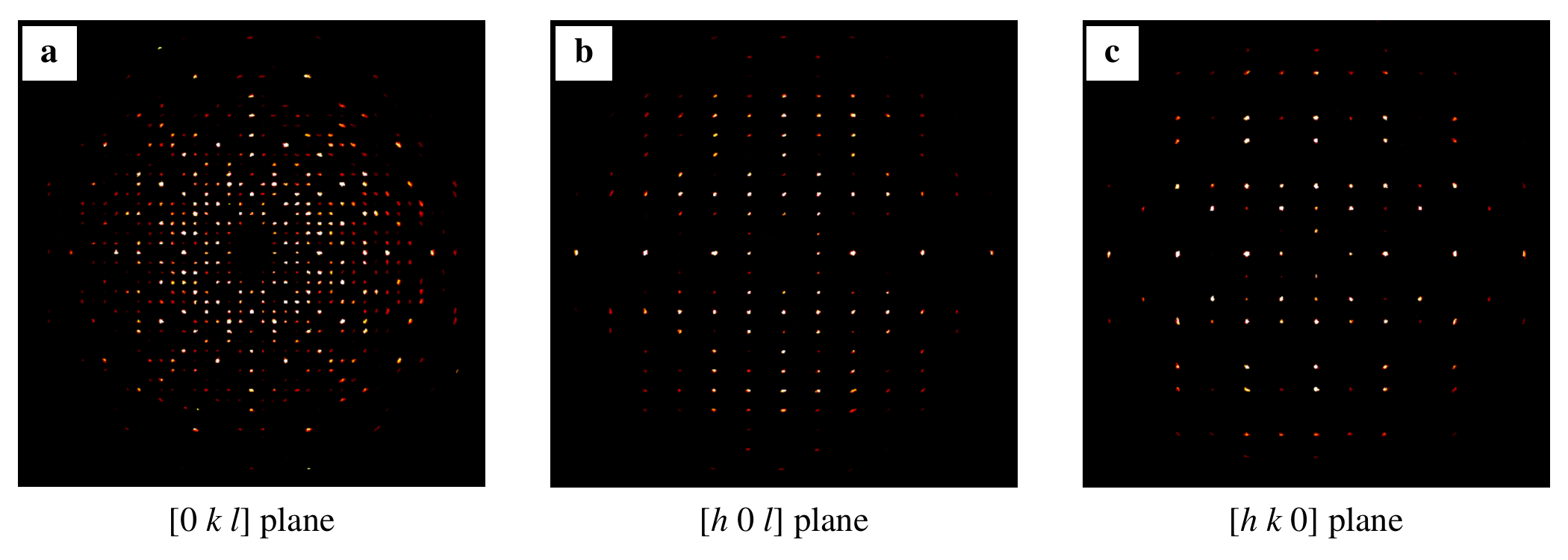}
	\label{Fig.S3}
\vspace*{-30pt}
\small
\begin{flushleft}
\justifying{
    \textbf{Fig. S3 $|$ Single-crystal XRD patterns of CeCo$_2$Ga$_8$.} \textbf{a} [0 $k$ $l$] plane. \textbf{b} [$h$ 0 $l$] plane. \textbf{c} [$h$ $k$ 0] plane. These patterns are neat without discernible impurity phase.
    }
\end{flushleft}
\normalsize
\end{figure*}

\begin{table}[!ht]
\tabcolsep 0pt \caption{\label{tbl:Structure} Single-crystal XRD refinement for CeCo$_2$Ga$_8$ at 273 K. Measurements with Mo $K_\alpha$ (0.71073 \AA). Refinement software: SHELXL-2018/3.}
\vspace*{-25pt}
\begin{center}
\def\temptablewidth{1\columnwidth}
{\rule{\temptablewidth}{1pt}}
\begin{tabular*}{\temptablewidth}{@{\extracolsep{\fill}}cc}
Composition ~~~~~~~~~~~~~~~~~~~~~~~~~~~~~~~~~~~         &  CeCo$_2$Ga$_8$           \\
Formula weight (g/mol) ~~~~~~~~~~~~~~~~~~~~      &  815.74                      \\
Crystal system ~~~~~~~~~~~~~~~~~~~~~~~~~~~~~~~~      & Orthorhombic                 \\
Space group ~~~~~~~~~~~~~~~~~~~~~~~~~~~~~~~~~~~         &  Pbam (No. 55)             \\
$a$ (\AA) ~~~~~~~~~~~~~~~~~~~~~~~~~~~~~~~~~~~~~~~~~~~           &  12.404(3)     \\
$b$ (\AA) ~~~~~~~~~~~~~~~~~~~~~~~~~~~~~~~~~~~~~~~~~~~           &  14.261(3)      \\
$c$ (\AA)  ~~~~~~~~~~~~~~~~~~~~~~~~~~~~~~~~~~~~~~~~~~~          &  4.0455(8)   \\
$V$ (\AA$^{3}$) ~~~~~~~~~~~~~~~~~~~~~~~~~~~~~~~~~~~~~~~~~~     &  715.6(3)    \\
$Z$  ~~~~~~~~~~~~~~~~~~~~~~~~~~~~~~~~~~~~~~~~~~~~~~~~~ &  4               \\
$\rho$ (g/cm$^{3}$) ~~~~~~~~~~~~~~~~~~~~~~~~~~~~~~~~~~~~~~ &  7.572                     \\
$2\theta$ range ~~~~~~~~~~~~~~~~~~~~~~~~~~~~~~~~~~~~~~~~     &  6.572 - 64.477$^{\circ}$         \\
$h, k, l$ range ~~~~~~~~~~~~~~~~~~~~~~~~~~~~~~~~~~~~ & $-19\leq h\leq 19$, $-21\leq k\leq 21$, $-6\leq l\leq 6$  \\
No. of reflections ~~~~~~~~~~~~~~~~~~~~~~~~~~~~~  &   16341                           \\
No. of independent reflections ~~~~~~~~~~~~ &   1538                                   \\
$R_1^{\dag}$, $w R_2^{\ddag}$ [$I>2\delta(I)$] ~~~~~~~~~~~~~~~~~~~~~~~~~~~&   2.17\%, 6.30\%    \\
\end{tabular*}
{\rule{\temptablewidth}{1pt}}
\end{center}
\small
\vspace*{-30pt}
\begin{flushleft}
\justifying{
$^{\dag}$ $R_1=\sum||F_{obs}|-|F_{cal}||/\sum|F_{obs}|$.\\
$^{\ddag}$ $w R_2=[\sum w(F_{obs}^2-F_{cal}^2)^2/\sum w (F_{obs}^2)^2]^{1/2}$, \\
$~~~~w=1/[\sigma^2F_{obs}^2+(a\cdot P)^2+b\cdot P]$, where  $P=[\mathrm{max}(F_{obs}^2)+2F_{cal}^2]/3$.
}
\end{flushleft}
\normalsize
\end{table}

\begin{table}[!ht]
\tabcolsep 0pt \caption{\label{tb2:Atoms} Atomic coordinates and equivalent isotropic displacement parameters $U_{eq}$ of CeCo$_2$Ga$_8$. $U_{eq}$ is taken as 1/3 of the trace of the orthogonalized $U_{ij}$ tensor.}
\vspace*{-25pt}
\begin{center}
\def\temptablewidth{1\columnwidth}
{\rule{\temptablewidth}{1pt}}
\begin{tabular*}{\temptablewidth}{@{\extracolsep{\fill}}ccccccc}
Atoms   &  Wyck.  &  $x$          &     $y$      &     $z$      &     Occ.      & $U_{eq}$ (\AA$^2$) \\\hline
  Ce     &   4h    &  0.33918(2)   &  0.68147(2)  &    0.50000    &    1.000      &   0.0105(1)            \\
  Co1    &   4h    &  0.53463(4)   &  0.90560(4)  &    0.50000    &    1.000      &   0.0085(1)            \\
  Co2    &   4h    &  0.65370(4)   &  0.59537(4)  &    0.50000    &    1.000      &   0.0086(1)            \\
  Ga1    &   2c    &  0.44947(3)   &  0.81835(2)  &    0.00000    &    1.000      &   0.0108(1)            \\
  Ga2    &   4g    &  0.73523(4)   &  0.67433(3)  &    0.00000    &    1.000      &   0.0107(1)            \\
  Ga3    &   4g    &  0.52119(5)   &  0.63219(5)  &    0.00000    &    1.000      &   0.0109(1)            \\
  Ga4    &   4h    &  0.59298(2)   &  0.75102(2)  &    0.50000    &    1.000      &   0.0142(1)            \\
  Ga5    &   4g    &  0.66190(4)   &  0.87868(2)  &    0.00000    &    1.000      &   0.0105(1)            \\
  Ga6    &   4g    &  0.66664(3)   &  0.48925(3)  &    0.00000    &    1.000      &   0.0115(1)            \\
  Ga7    &   2b    &  0.83927(4)   &  0.54432(2)  &    0.50000    &    1.000      &   0.0117(1)            \\
  Ga8    &   4g    &  0.50000      &  0.50000     &    0.50000    &    1.000      &   0.0212(2)            \\
  Ga9    &   4h    &  0.50000      &  0.00000     &    0.00000    &    1.000      &   0.0127(1)            \\
\end{tabular*}
{\rule{\temptablewidth}{1pt}}
\end{center}
\end{table}

\textbf{SI \Rmnum{2}: Resistivity of CeCo$_2$Ga$_8$}\\

\begin{figure*}[!htp]
	\vspace*{-26pt}
	\includegraphics[width=10cm]{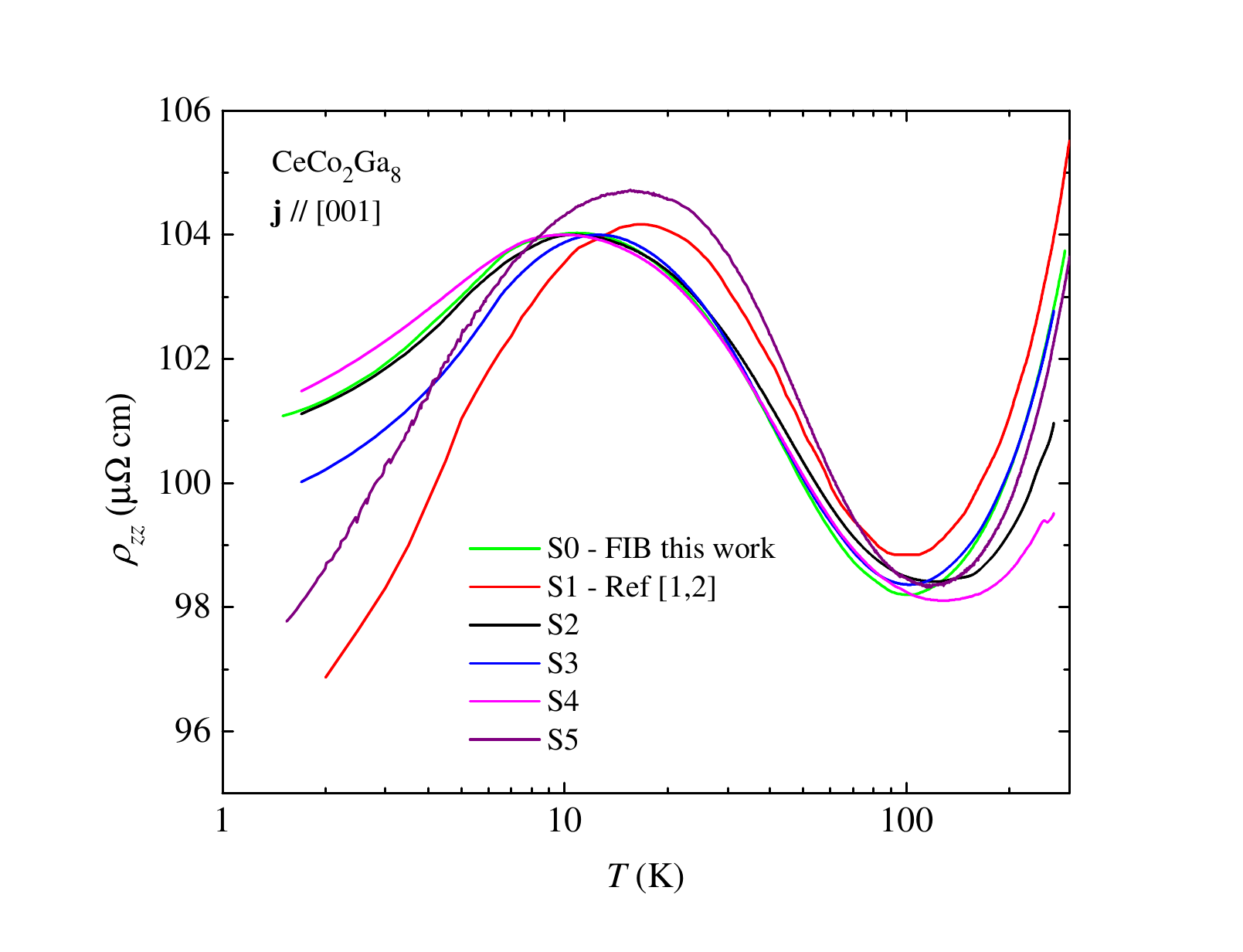}
	\label{Fig.S4}
\vspace*{-20pt}
\small
\begin{flushleft}
\justifying{
    \textbf{Fig. S4 $|$ Temperature dependent resistivity of CeCo$_2$Ga$_8$ with current along [001].} Data from various samples S0-S5. S0 is the FIB-microstructured device investigated in the main text of this work; the curve S1 is reproduced from our previous publications Refs.~\cite{SWangL-CeCo2Ga8,SChengK-CeCo2Ga8Q1D}; all S1-S5 are from standard four-probe measurements on bulk crystals.
    }
\end{flushleft}
\normalsize
\end{figure*}

The temperature dependent resistivity of CeCo$_2$Ga$_8$ is further studied on different samples S2-S5 in a standard four-probe measurement on bulk crystals, and the results are summarized in Fig.~S2. For comparison, we also show the curve reproduced from our previous publications \cite{SWangL-CeCo2Ga8,SChengK-CeCo2Ga8Q1D} (labeled by S1), and the FIB-microstructured device in the main text of this work (S0). Three prominent features can be identified: (\rmnum{1}) the RRRs of all the samples are commonly small, typically close to 1; (\rmnum{2}) Kondo coherence scale $T^*$ varies from sample to sample, in the range 10-20 K, and the $T^*\approx12$ K of the FIB sample falls within this range; (\rmnum{3}) FIB process seems to have no essential influence on $T^*$. Usually, small RRR means poor sample quality. However, for CeCo$_2$Ga$_8$, on the one hand, Laue XRD suggests decent crystallization; on the other hand, according to our experience, it seems that the RRR can hardly be improved. Besides the incoherent Kondo effect that has been discussed in the main text, another reason might be the large unit cell and the complicated crystalline structure with 12 different chemical sites, which potentially enhances the extent of ``disorder". The latter can be partly evidenced by the fact that the RRRs of most known $ReTm_2X_8$ ($Re=$ La, Ce, Pr, Nd, Sm; $Tm=$Fe, Co; $X=$ Al, Ga) compounds are generally less than 3 \cite{SITamura-LaFe2Al8,SXTDeng-CeFe2Ga8,SGosh-CeFe2Al8,SChengK-CeCo2Ga8Q1D,SOOMichael-PrFe2Ga8,SHSNair-PrFe2Al8,SJJXiao-PrCo2Al8,SOOMichael-PrCo2Ga8,SCXWang-NdFe2Ga8,SWHe-NdCo2Al8}. For CeCo$_2$Ga$_8$, because it sits nearby a quantum critical point \cite{SWangL-CeCo2Ga8}, its RRR can be further degraded due to the enhancement of residual resistivity. \\

\textbf{SI \Rmnum{3}: Additional PHE results for below $T^*$}\\

\begin{figure*}[!htp]
	\vspace*{-10pt}
	\includegraphics[width=10cm]{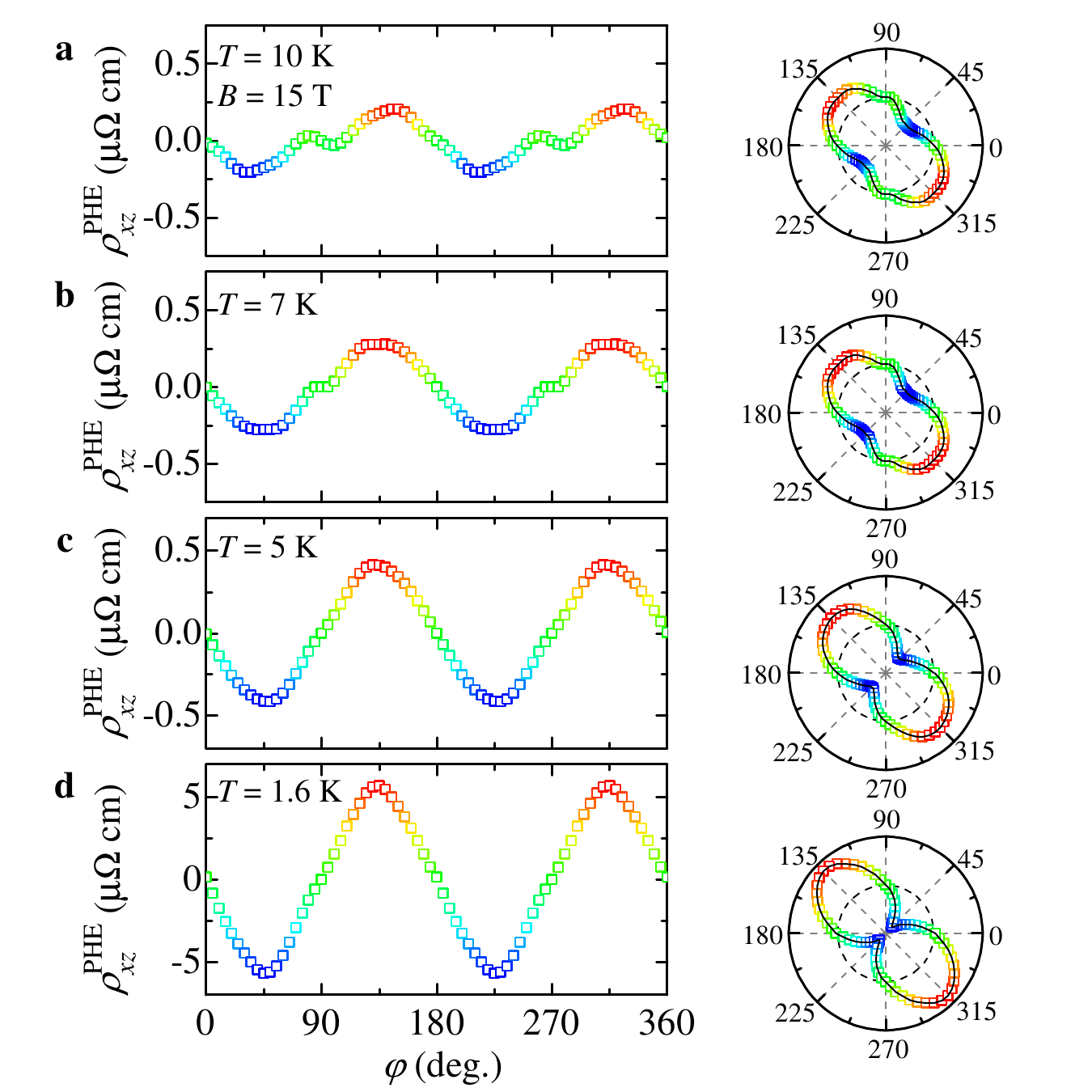}
	\label{Fig.S5}
\vspace*{-10pt}
\small
\begin{flushleft}
\justifying{
    \textbf{Fig. S5 $|$ More PHE data of CeCo$_2$Ga$_8$ below $T^*$.} The shoulder near 90$^\circ$ gradually diminishes, and meanwhile, the amplitude of PHE increases as $T$ decreases.
    }
\end{flushleft}
\normalsize
\end{figure*}

\newpage

\textbf{SI \Rmnum{4}: Numerical simulations of PHE}\\

In this section, we present the details of numerical simulations of PHE based on the hypothesis of dynamic hybridization. Following the similar treatment described in Ref.~\cite{SZhongJ-CPB2023}, in the magnetic field relevant coordinate, we define the electrical conductivity parallel with and perpendicular to the magnetic field as $\sigma^{c,h}_{\parallel}$ and $\sigma^{c,h}_{\perp}$, respectively, where the superscripts $c$ and $h$ respectively stand for the transport channels of non-hybridized conduction electrons and heavy-fermion quasiparticles. For the measurement coordinate relative to the electric current, a transformation from the magnetic-field-relevant coordinate to the electric-current-relevant coordinate is needed, and this is realized by a 2D rotation matrix $\hat{\mathbf{T}}$ whose form will be given below. After this transformation, the conductivities are:
\begin{equation}
\begin{aligned}
\boldsymbol{\sigma}^{c,h}=\hat{\mathbf{T}}\left[
  \begin{array}{cc}
    \sigma^{c,h}_{\parallel} & 0                          \\
    0                        & \sigma^{c,h}_{\perp}       \\
  \end{array}
\right]\hat{\mathbf{T}}^{\dag}.
\end{aligned}
\end{equation}

According to the idea of two-fluid model, we introduce $w$, the weight of conductivity through hybridized heavy-mass quasiparticles, thereby the weight contributed from non-hybridized conduction electrons is $1-w$. The total electrical conductivity tensor $\boldsymbol{\sigma}$ can be given by:
\begin{equation}
\begin{aligned}
\boldsymbol{\sigma}=(1-w)\boldsymbol{\sigma}^{c}+w\boldsymbol{\sigma}^{h}=(1-w)\hat{\mathbf{T}}_c\left[
  \begin{array}{cc}
    \sigma^{c}_{\parallel} & 0                          \\
    0                        & \sigma^{c}_{\perp}       \\
  \end{array}
\right]\hat{\mathbf{T}}_c^{\dag}+w\hat{\mathbf{T}}_h\left[
  \begin{array}{cc}
    \sigma^{h}_{\parallel} & 0                          \\
    0                        & \sigma^{h}_{\perp}       \\
  \end{array}
\right]\hat{\mathbf{T}}_h^{\dag}.
\end{aligned}
\end{equation}
For temperature above $T_K^{on}$, the $f$ electrons are fully localized, and only conduction-electron Fermi surface from Co/Ga orbitals are present along $\Gamma$-$\text{Y}$ whose rotation matrix about $\mathbf{b}$ axis is approximately:
\begin{equation}
\begin{aligned}
\hat{\mathbf{T}}_c=\left[
  \begin{array}{cc}
    \cos{\varphi}  & -\sin{\varphi}      \\
    \sin{\varphi}  & \cos{\varphi}       \\
  \end{array}
\right]~~~~~~~(T>T_K^{on}).
\end{aligned}
\end{equation}
For $T^*<T<T_K^{on}$, due to the incoherent Kondo scattering and the precursor fluctuating hybridization, growing higher harmonics start to play a role. For simplicity, we only consider a single $A_4$ mode which is expressed by $a_4 \equiv A_4/A_2$. This can be realized by adding a new term with higher-order rotation matrix \cite{SPRout-KaAlO3PHE}:
\begin{equation}
\begin{aligned}
\hat{\mathbf{T}}_c^{(4)}=\left[
  \begin{array}{cc}
    \cos{2\varphi}  & -\sin{2\varphi}      \\
    \sin{2\varphi}  & \cos{2\varphi}       \\
  \end{array}
\right]~~~~~~~(T^*<T<T_K^{on}).
\end{aligned}
\end{equation}
For temperature below $T^*$, new heavy-electron Fermi sheet comes into being, and its rotation matrix $\hat{\mathbf{T}}_h$ should take the same form as in Eq.~(S3).

With these assumptions, we derive the expressions for the conductivity of PAMR and PHE,
\begin{equation}
\begin{aligned}
\sigma_{zz}=\left\{
  \begin{array}{ll}
    \sigma_{\parallel}^c-\Delta\sigma^c\sin^2{\varphi}  &~~(T>T_K^{on})         \\
    \sigma_{\parallel}^c-\Delta\sigma^c(\sin^2{\varphi}+a_4\sin^2{2\varphi})                &~~(T^*<T<T_K^{on})                \\
    (1-w)[\sigma_{\parallel}^c-\Delta\sigma^c(\sin^2{\varphi}+a_4\sin^2{2\varphi})]+w(\sigma_{\parallel}^h-\Delta\sigma^h\sin^2{\varphi})   &~~(T<T^*)
   \end{array}
\right.
\end{aligned}
\end{equation}
\begin{equation}
\begin{aligned}
\sigma_{zx}=\left\{
  \begin{array}{ll}
   \Delta\sigma^c \sin{\varphi}\cos{\varphi}  &~~(T>T_K^{on}) \\
    \Delta\sigma^c\left(\sin{\varphi}\cos{\varphi}+a_4\sin{2\varphi}\cos{2\varphi}\right)                &~~(T^*<T<T_K^{on})                \\
    (1-w)    \Delta\sigma^c\left(\sin{\varphi}\cos{\varphi}+a_4\sin{2\varphi}\cos{2\varphi}\right)  +w    \Delta\sigma^h\sin{\varphi}\cos{\varphi}       &~~(T<T^*)     \\
   \end{array}
\right.
\end{aligned}
\end{equation}
Here, $\Delta \sigma^{c,h}\equiv\sigma_{\parallel}^{c,h}-\sigma_{\perp}^{c,h}$. Finally, the conductivity tensor is converted back into resistivity tensor via $\boldsymbol{\rho}=\boldsymbol{\sigma}^{-1}$ for comparison with the measurable values, and the off-diagonal component of $\boldsymbol{\rho}$ yields $\rho_{xz}^\text{PHE}$.

In the numerical simulation, we ignored the temperature dependence of conductivities. $\sigma_{\parallel}^c=1.00$, $\sigma_{\perp}^c=0.90$, $\sigma_{\parallel}^h=0.16$, and $\sigma_{\perp}^h=0.20$ are used, which appear reasonable in accordance to our experimental results. Temperature dependent weight $w(T)$ and $a_4$ are exploited to reproduce the evolution of the two conducting channels. The calculated PHE patterns are displayed in Fig.~5 of the main text: (a) $w$=0.0 and $a_4$=0.0 for $T>T_K^{on}$; (b) $w$=0.0 and $a_4$=$-1.0$ for $T^*<T<T_K^{on}$; (c) $w$=0.7 and $a_4$=$-0.3$ for $T<T^*$; (d) $w$=1.0 and $a_4$=0.0 for well below $T^*$.\\

\textbf{SI \Rmnum{5}: Hall effect of CeCo$_2$Ga$_8$}\\

\begin{figure*}[!htp]
	\vspace*{-20pt}
	\includegraphics[width=16cm]{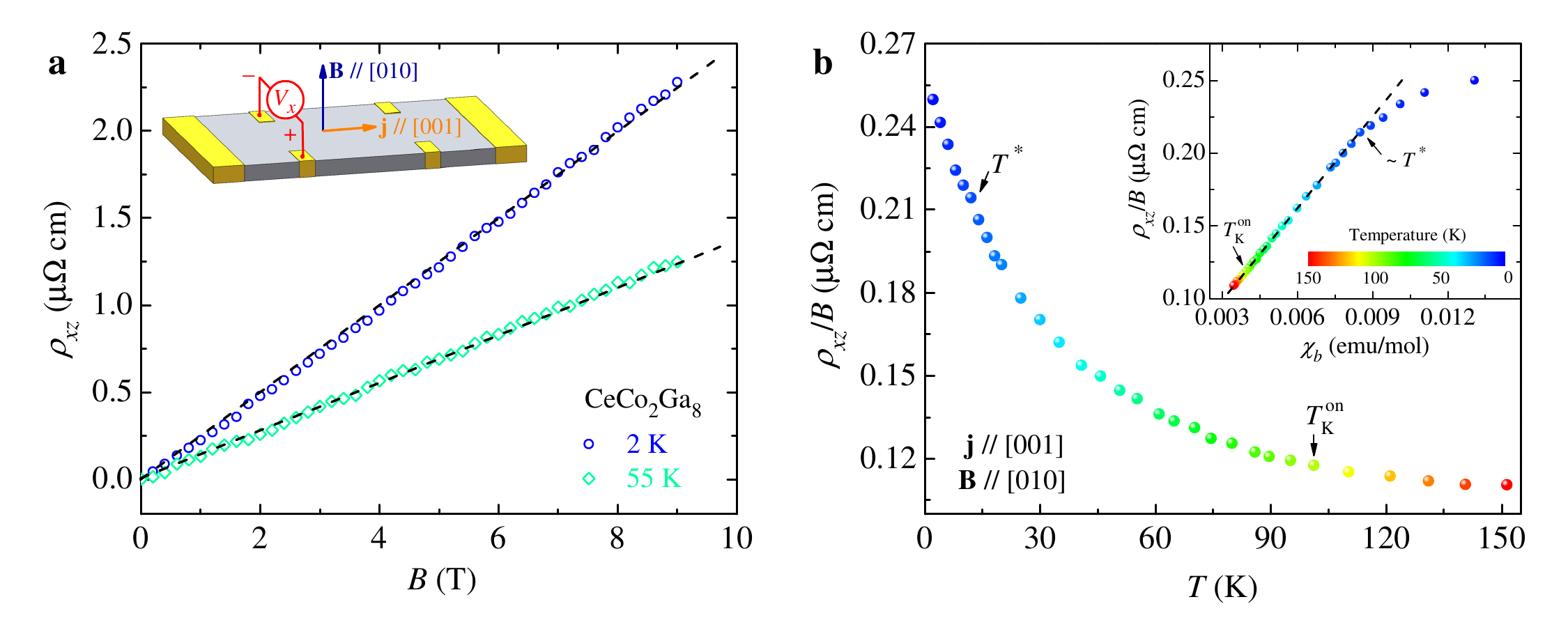}
	\label{Fig.S6}
\vspace*{-15pt}
\small
\begin{flushleft}
\justifying{
    \textbf{Fig. S6 $|$ Hall effect of CeCo$_2$Ga$_8$.} \textbf{a} Isothermal field dependent Hall resistivity at 2 and 55 K; the inset shows the configuration of Hall effect measurements: electrical current $\mathbf{j}\parallel[001]$, and $\mathbf{B}\parallel[010]\perp\mathbf{j}$. \textbf{b} $\rho_{xz}/B$ as a function of $T$; the inset shows a plot of $\rho_{xz}/B$ vs. $\chi_b$ with temperature as an implicit parameter.
    }
\end{flushleft}
\normalsize
\end{figure*}

Hall effect with a normal configuration ($\mathbf{j}\parallel[001]$, and $\mathbf{B}\parallel[010]$) was also measured on CeCo$_2$Ga$_8$, the main results of which are summarized in Fig.~S4. Over the full temperature range 2-150 K, the Hall resistivity $\rho_{xz}$ ($\equiv E_x/j_z$) is linear with $B$, as exemplified by the data sets at 2 K and 55 K. However, linear $\rho_{xz}(B)$ does NOT necessarily mean the absence of anomalous Hall effect (AHE) \cite{SHeX-Ce3TiSb5}. A tentative plot of $\rho_{xz}/B$ vs. $T$ clearly shows a Curie-Weiss-like behavior, reminiscent of the involvement of the AHE component $\rho_{xz}^{\text{AHE}}$ that is proportional to magnetization. Apparently, AHE dominates over the ordinary Hall effect, which prevents us from extracting the carrier density and mobility from Hall effect. However, qualitatively, the carrier mobility should be rather low considering the linear $\rho_{xz}(B)$, in consistency with the small RRR. Historically, the mechanism of the AHE in heavy-fermion compounds is usually ascribed to skew scatterings. We here refer to our recent work \cite{SHeX-Ce3TiSb5} and the relevant references therein.

An alternative plot of $\rho_{xz}/B$ vs $\chi_b$ with $T$ an implicit parameter is then given in the inset to Fig.~S4(b). Indeed, a linear scaling is found between $\rho_{xz}/B$ and $\chi_b$, and such linear scaling persists until $T^*$ where it is gradually violated. In other words, $\rho_{xz}$ appears essentially featureless in the intermediate regime between $T_K^{on}$ and $T^*$, in sharp contrast to PHE. This implies that classic Hall effect is not sensitive to the dynamic hybridization regime.

\end{document}